\documentclass[aps,11pt,
nofootinbib,groupedaddress,superscriptaddress]{revtex4-2}

\usepackage{amssymb, amsfonts, amsmath, bm}
\usepackage[
]{graphicx}
\usepackage{color}
\usepackage{mathrsfs}

\usepackage[
]{hyperref}
\hypersetup{%
 colorlinks=true,
  linkcolor=blue,   
  citecolor=blue,   
  urlcolor=blue     
}
\newcommand{\orcid}[1]{%
  \href{https://orcid.org/#1}{\includegraphics[height=0.7em]{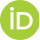}}%
}

\begin{document}
\title{Deflection angle in the strong deflection limit for static and axisymmetric spacetimes:
Local curvature, matter fields, and quasinormal modes}
\author{Takahisa Igata\:\!\orcid{0000-0002-3344-9045}
}
\email{takahisa.igata@gakushuin.ac.jp}
\affiliation{
Department of Physics, Gakushuin University,\\
Mejiro, Toshima, Tokyo 171-8588, Japan}
\date{\today}

\begin{abstract}
We investigate the deflection of photons in the strong deflection limit within 
static and axisymmetric
spacetimes possessing reflection symmetry. As the impact parameter approaches its critical value, the deflection angle exhibits a logarithmic divergence. This divergence is characterized by a logarithmic coefficient and a constant offset, which we express in terms of the coordinate-invariant curvature quantities evaluated at the unstable 
circular photon orbit.
The curvature contribution is encoded in the electric part of the Weyl tensor, reflecting tidal effects, and the matter contribution is encoded in the Einstein tensor, capturing the influence of local energy and pressure. We also express these coefficients using 
the 
Newman--Penrose scalars. By exploiting the relationship between the strong deflection limit and 
the quasinormal modes, we derive a new expression for the quasinormal mode frequency in the eikonal limit in terms of the curvature scalars. Our results provide a unified and coordinate-invariant framework that connects observable lensing features and quasinormal modes to the local geometry and matter distribution near compact objects.
\end{abstract}

\maketitle

\section{Introduction}
\label{sec:1}
Recent observational advances, such as the imaging of black hole shadows by the Event Horizon Telescope~\cite{EventHorizonTelescope:2019dse,EventHorizonTelescope:2022wkp} and the detection of gravitational waves by LIGO and Virgo~\cite{LIGOScientific:2016aoc}, have significantly enhanced our capacity to explore spacetime geometry in the strong-field regime.
In particular, the imaging of black hole shadows has revealed how light propagates near compact objects, through the observed emission from the surrounding accretion flow.
At the same time, gravitational lensing has long provided a theoretical basis for understanding the motion of photons through curved spacetimes~(see, e.g., Ref.~\cite{Perlick:2004} for a review).

When light rays pass extremely close to a compact object, the deflection angle becomes large. In this regime, the strong deflection limit (SDL) provides a useful approximation for analyzing gravitational lensing. The deflection angle diverges logarithmically as the impact parameter approaches its critical value, with a characteristic rate and a constant offset, known as the SDL coefficients. A systematic description
of this behavior was developed by Bozza~\cite{Bozza:2002zj}, and has since been refined and extended in several works~\cite{Tsukamoto:2016jzh,Shaikh:2019itn}. 
These methods have been widely applied to various static and spherically symmetric spacetimes, including 
the Schwarzschild spacetime~\cite{Bozza:2001xd}, 
the Reissner--Nordstr\"om spacetime~\cite{Eiroa:2002mk,Tsukamoto:2016oca}, 
regular black hole spacetime~\cite{Eiroa:2010wm}, 
naked singularity spacetime~\cite{Chen:2023uuy}, 
wormhole spacetimes~\cite{TejeiroS:2005ltc,Nandi:2006ds,Tsukamoto:2016qro,Nandi:2016uzg,Shaikh:2019jfr,Bhattacharya:2019kkb,Izmailov:2019tyq}, gravastar spacetimes~\cite{Kubo:2016ada}, higher-dimensional black hole spacetimes~\cite{Chakraborty:2016lxo}, and black hole spacetimes with quantum corrections~\cite{Soares:2024rhp,Nascimento:2020ime}. Bozza's method has also inspired several extensions, such as finite-distance corrections~\cite{Ishihara:2016sfv,Takizawa:2021gdp}, the case of massive particles~\cite{Feleppa:2024kio}, different source configurations~\cite{Gao:2025arj}, and marginally unstable photon spheres~\cite{Sasaki:2025web}. Furthermore, the same formulation applies to the reflection-symmetric plane of axisymmetric spacetimes, including the Kerr~\cite{Bozza:2002af,Dolan:2010wr}, the Kerr--Newman~\cite{Hsieh:2021scb,AbhishekChowdhuri:2023ekr}, the Majumdar--Papapetrou~\cite{Patil:2016oav}, and the Zipoy--Voorhees spacetimes~\cite{Chakrabarty:2022fbd}.

Recently, it was shown that, in 
static and spherically symmetric spacetimes, the SDL coefficients can be expressed in terms of local geometric and matter field quantities at the photon sphere~\cite{Igata:2025taz}. This coordinate-independent formulation refines earlier coordinate-based approaches and provides a framework that links local curvature and matter distribution to observable features of strong gravitational lensing. However, whether a similar local and coordinate-independent description remains valid in less symmetric spacetimes has not yet been fully clarified.

Independent of these developments, quasinormal mode (QNM) frequencies have also been found to reflect the properties of unstable circular photon orbits in the eikonal limit. Ferrari and Mashhoon~\cite{Ferrari:1984zz} first established the correspondence between QNM frequencies and the orbital parameters of the unstable circular photon orbit in this limit. Subsequently, Cardoso \textit{et al.}~\cite{Cardoso:2008bp} generalized this relation to a wide class of static and stationary spacetimes, showing that the imaginary part of the QNM frequency is determined by the Lyapunov exponent of the unstable photon orbit. 
Stefanov \textit{et al.}~\cite{Stefanov:2010xz} first pointed out a quantitative connection between the SDL and QNMs, 
showing that the imaginary part of the QNM frequency in the eikonal limit is inversely proportional to the logarithmic coefficient in the SDL. 
Raffaelli~\cite{Raffaelli:2014ola} further clarified the geometrical interpretation of this correspondence and discussed its implications for wave propagation near compact objects.
More significantly, recent work~\cite{Igata:2025taz} has shown that, in static and spherically symmetric spacetimes, the QNM frequency in the eikonal limit can be expressed in terms of local geometric and matter field quantities evaluated at the photon sphere. This coordinate-invariant formulation provides a physical insight into the QNM frequency as a manifestation of both spacetime curvature and matter distribution near the unstable photon orbit, thereby offering a unified geometric perspective complementary to that of the strong deflection limit.

The relation between QNMs and the SDL is particularly significant in light of recent gravitational wave observations, where the ringdown phase has been detected 
and found to be consistent, within the current observational uncertainties, with the theoretical predictions of general relativity~\cite{LIGOScientific:2016aoc,LIGOScientific:2016dsl,LIGOScientific:2020tif, Bolokhov:2025uxz}. This consistency between the measured and predicted QNM frequencies provides empirical support for the validity of the underlying theoretical framework and motivates further efforts to connect these modes with other observable phenomena in strong gravitational fields. Since the QNM frequencies in the eikonal limit are determined by the properties of unstable photon orbits, 
this underscores the importance of understanding the local geometry near the photon sphere in interpreting strong-field gravitational wave signals.

In this work, we generalize the coordinate-invariant formulation of the deflection angle in the SDL to static and axisymmetric spacetimes with reflection symmetry. We show that the SDL coefficients can be expressed in terms of local geometric and matter field quantities evaluated at the unstable 
circular photon orbit.
Furthermore, we reformulate these coefficients using curvature-based quantities, such as the electric part of the Weyl tensor and Newman--Penrose (NP) scalars and apply the formalism to several explicit spacetimes to illustrate its physical relevance. Using the relation between the SDL and the QNM, we derive a new expression for the QNM frequency in terms of local curvature and matter field quantities. 

This paper is organized as follows. In Sec.~\ref{sec:2}, we introduce the general formalism of photon dynamics in 
static and axisymmetric
spacetimes and derive the conditions for the existence of circular photon orbits. In Sec.~\ref{sec:3}, we derive the deflection angle in the SDL by isolating the logarithmically divergent contribution from the photon trajectory integral. In Sec.~\ref{sec:4}, we recast the SDL expression in a coordinate-invariant form, demonstrating that it depends only on the circumferential radius of the unstable 
circular photon orbit
and local curvature. In Sec.~\ref{sec:5}, we relate this expression to the matter field quantities. In Sec.~\ref{sec:6}, we further recast the deflection angle in terms of the NP scalars. In Sec.~\ref{sec:7}, we present several applications of the formalism developed herein. In Sec.~\ref{sec:8}, we establish the correspondence between QNM frequencies and SDL coefficients in 
static and axisymmetric
spacetimes. Finally, in Sec.~\ref{sec:9}, we summarize our findings and discuss their implications for gravitational lensing in the SDL. 

Throughout this paper, we employ the abstract index notation~\cite{Wald:1984} and use geometrized units in which the gravitational constant and the speed of light are set to unity.

\section{Unstable circular photon orbits in static and axisymmetric spacetimes}
\label{sec:2}
We consider a general 
static and axisymmetric
spacetime, which admits Killing fields corresponding to time translations and rotations, adapted to the 
time
coordinate $t$ and 
the azimuthal coordinate
$\varphi$, respectively. The canonical form of the line element is given by
\begin{align}
\mathrm{d}s^2 = -e^{2\:\!\psi}\,\mathrm{d}t^2
+ e^{-2\:\!\psi}\left[\:\!
e^{2\:\!\gamma}\left(\mathrm{d}\rho^2+\mathrm{d}\zeta^2\right)
+W^2\:\!\mathrm{d}\varphi^2\:\!\right],
\label{eq:metric}
\end{align}
where $\psi$, $\gamma$, and $W$ are functions of 
the spatial coordinates $\rho$ and $\zeta$ in the meridional plane, which can be regarded as cylindrical \(\rho\)-\(z\)-type coordinates in a particular gauge~\cite{Griffiths:2012}. 
We assume a reflection symmetry (i.e., a $\mathbb{Z}_2$ symmetry) with respect to the $\zeta=0$ plane, which requires that $\psi$, $\gamma$, and $W$ are even functions of $\zeta$:
\begin{align}
\psi(\rho,-\zeta)=\psi(\rho,\zeta), \quad \gamma(\rho,-\zeta)=\gamma(\rho,\zeta), \quad W(\rho,-\zeta)=W(\rho,\zeta).
\label{eq:Z2}
\end{align}
We define the auxiliary function
\begin{align}
R(\rho, \zeta)=e^{-\psi} W,
\label{eq:R}
\end{align}
which represents the circumferential radius around the symmetry axis ($\rho=0$). We assume $\rho\in (0, \rho_{\infty})$ and that $R$ diverges as $\rho \to \rho_{\infty}$ for any fixed $\zeta$. 
Furthermore, on the $\zeta=0$ plane, we assume that the spacetime is asymptotically locally flat; that is, as $\rho \to \rho_{\infty}$, the metric functions approach 
the Minkowski values: $e^{2\psi} \to 1$, $e^{2\gamma} \to 1$, and $W \to \infty$.
Note that asymptotic flatness assumed only on the equatorial plane does not guarantee asymptotic flatness off the plane. However, since the following analysis considers the deflection angle of null geodesics confined to the equatorial plane, this assumption is sufficient for our purpose.

We assume that photons follow null geodesics. Restricting our analysis to the reflection-symmetric plane $\zeta=0$, where every initially confined photon remains there, we obtain the reduced Lagrangian
\begin{align}
\mathscr{L}=\frac{1}{2}\left[\:\!
-e^{2\:\!\psi} \:\!\dot{t}^2+e^{-2\:\!\psi}\left(
e^{2\:\!\gamma}\:\!\dot{\rho}^2+W^2 \dot{\varphi}^2
\right)
\:\!\right],
\end{align}
where $\psi$, $\gamma$, and $W$ are evaluated at $\zeta=0$, and the overdot denotes differentiation with respect to an affine parameter along the null geodesic. Since $\mathscr{L}$ is independent of $t$ and $\varphi$, the corresponding conjugate energy and angular momentum, 
\begin{align}
E&=e^{2\:\!\psi}\:\! \dot{t},
\label{eq:E}
\\
L&=e^{-2\:\!\psi} W^2 \dot{\varphi},
\label{eq:L}
\end{align}
respectively, are conserved. 

By combining the null condition $\mathscr{L}=0$ with Eqs.~\eqref{eq:E} and \eqref{eq:L}, we obtain
\begin{align}
\dot{\rho}^2+e^{-2\:\!\gamma} \left(
\frac{e^{4\:\!\psi}}{W^2}L^2-E^2
\right)=0.
\label{eq:nullcond}
\end{align}
Assuming that 
$\dot{\varphi}\neq 0$,
we divide Eq.~\eqref{eq:nullcond} by $\dot{\varphi}^2$ to yield the radial orbital differential equation:
\begin{align}
\bigg(\frac{\mathrm{d}\rho}{\mathrm{d}\varphi}\bigg)^2+V(\rho)=0,
\end{align}
where the effective potential $V(\rho)$ is defined as
\begin{align}
V(\rho)=e^{-2\:\!\gamma} W^2\left(1-\frac{e^{-4\:\!\psi} W^2}{b^2}\right),
\label{eq:V}
\end{align}
with the impact parameter defined as $b\equiv L/E$.

Next, we consider circular orbits, for which $\dot{\rho}=0$ and $\ddot{\rho}=0$. Consequently, the effective potential $V$ and its derivative $V'$ vanish at $\rho=\rho_{\mathrm{m}}$, where $\rho_{\mathrm{m}}$ denotes the 
radius coordinate
of the circular orbit. Setting $V(\rho_{\mathrm{m}})=0$ immediately yields the critical impact parameter $b=b_{\mathrm{c}}$ with 
\begin{align}
b_{\mathrm{c}}\equiv e^{-2\:\!\psi_{\mathrm{m}}} W_{\mathrm{m}},
\label{eq:bc}
\end{align}
where, hereafter, the subscript $\mathrm{m}$ indicates evaluation at $\rho=\rho_{\mathrm{m}}$ and $\zeta=0$ [e.g., $\psi_{\mathrm{m}}\equiv \psi(\rho_{\mathrm{m}},0)$ and $W_{\mathrm{m}}\equiv W(\rho_{\mathrm{m}},0)$]. Furthermore, imposing $V'(\rho_{\mathrm{m}})=0$ leads to
\begin{align}
\psi'_{\mathrm{m}}=\frac{W'_{\mathrm{m}}}{2W_{\mathrm{m}}}.
\label{eq:ccond2}
\end{align}
Evaluating $V''$ at $\rho=\rho_{\mathrm{m}}$ for $b=b_{\mathrm{c}}$, we obtain 
\begin{align}
V''_{\mathrm{m}}=2 e^{-2\gamma_{\mathrm{m}}}\left[\:\!
2W_{\mathrm{m}}^2 \psi''_{\mathrm{m}}+(W'_{\mathrm{m}})^2-W_{\mathrm{m}}W''_{\mathrm{m}}
\:\!\right],
\label{eq:V"m}
\end{align}
which provides a criterion for the stability of the circular orbits: the orbit is unstable if $V''_{\mathrm{m}}<0$, and stable if $V''_{\mathrm{m}}>0$. In what follows, we focus on the unstable 
circular photon orbits
(i.e., $V''_{\mathrm{m}}<0$), and hence, 
\begin{align}
2W_{\mathrm{m}}^2 \psi''_{\mathrm{m}}+(W'_{\mathrm{m}})^2-W_{\mathrm{m}}W''_{\mathrm{m}}<0.
\end{align}

\section{Deflection angle in the strong deflection limit}
\label{sec:3}
In this section, we derive the deflection angle in the SDL, following the approach in Ref.~\cite{Bozza:2002zj}. Let $\rho_0$ denote the radial coordinate of the closest approach, where $V(\rho_0)=0$. Then, the impact parameter is given by
\begin{align}
b=e^{-2\:\!\psi_0} W_0.
\label{eq:batrho0}
\end{align}
Throughout the manuscript, the subscript $0$ denotes evaluation at $\rho=\rho_{0}$ and $\zeta=0$ [e.g., 
$\psi_0\equiv \psi(\rho_0, 0)$ and $W_0\equiv W(\rho_0,0)$]. 
We define the integral
\begin{align}
I(\rho_0)=2 \int_{\rho_0}^{\rho_{\infty}} \frac{|\mathrm{d}\rho|}{\sqrt{-V}}
\end{align}
which represents the total angular change experienced by a photon traveling from infinity to $\rho_0$ and then back to infinity. 
Since the radial coordinate \(\rho\) may decrease as it approaches spatial infinity in the chosen coordinate system, the absolute value ensures that the integrand remains positive regardless of whether \(\rho\) increases or decreases relative to the turning point \(\rho_0\).
Finally, the deflection angle is defined by
\begin{align}
\alpha(\rho_0)=I(\rho_0)-\pi. 
\end{align}

To analyze the behavior of the deflection angle in the SDL, we introduce a new variable, as proposed in Ref.~\cite{Igata:2025taz}, defined by 
\begin{align}
z=1-\frac{R_0}{R},
\end{align}
where $R$ is defined in Eq.~\eqref{eq:R} as a function of $\rho$ and $\zeta$ and is evaluated on the $\zeta=0$ plane.
In terms of this variable, the integral $I(\rho_0)$ becomes 
\begin{align}
I(\rho_0)=2 \int_0^1 \frac{\mathrm{d}z}{\sqrt{-\frac{(R')^2}{R_0^2}(1-z)^4\:\! V}},
\label{eq:Iz}
\end{align}
where $R'$ denotes the derivative of $R(\rho,0)$ with respect to $\rho$. By expanding $(R')^2$ and $V$, as given in Eqs.~\eqref{eq:R} and \eqref{eq:V}, in powers of $z$, we obtain
\begin{align}
(R')^2&=(R'_0)^2+2R_0R_0'' z+O(z^2),
\\
V&= \frac{R_0 V'_0}{R_0'} \:\! z + \left[\:\!
\left(\frac{R_0}{R_0'}-\frac{R_0^2R_0''}{2\:\!(R_0')^3}\right)V_0'
+\frac{R_0^2 }{2\:\!(R_0')^2}V_0''
\:\!\right]z^2+ O(z^3),
\end{align}
where we have used Eq.~(17).
Hence, truncating the expression under the square root in Eq.~\eqref{eq:Iz} to second order in $z$ yields $\sqrt{c_1 z + c_2 z^2}$. Accordingly, we define the corresponding integral as 
\begin{align}
I_{\mathrm{D}}(\rho_0)=2\int_0^1 \frac{\mathrm{d}z}{\sqrt{c_1 z+c_2 z^2}},
\label{eq:ID}
\end{align}
where the coefficients are given by
\begin{align}
c_1&=-\frac{R'_0 V'_0}{R_0},
\label{eq:prec1}
\\[2mm]
c_2&=3\left(
\frac{R'_0V'_0}{R_0}-\frac{R''_0 V'_0}{2 R'_0}
\right)-\frac{V''_0}{2}.
\label{eq:prec2}
\end{align}

The integral $I_{\mathrm{D}}$ isolates the leading-order divergent behavior of the total integral $I$ in the SDL, i.e., as $\rho_0 \to \rho_{\mathrm{m}}$. Since $V'_0\to 0$ in this limit, the integrand in Eq.~\eqref{eq:Iz} reduces to $1/(\sqrt{c_2} z)$, which leads to a logarithmic divergence. Thus, we identify $I_{\mathrm{D}}$ as the divergent part of $I$, and define the regular part as
\begin{align}
I_{\mathrm{R}}(\rho_0)=I(\rho_0)-I_{\mathrm{D}}(\rho_0).
\end{align}
The regular part $I_{\mathrm{R}}$ typically may require numerical evaluation depending on the global structure of the spacetime. Evaluating the integral~\eqref{eq:ID} yields
\begin{align}
I_{\mathrm{D}}(\rho_0)=\frac{4}{\sqrt{c_2}} \log \frac{\sqrt{c_1+c_2}+\sqrt{c_2}}{\sqrt{c_1}}. 
\label{eq:IDc}
\end{align}

To express the SDL (i.e., $\rho_0 \to \rho_{\mathrm{m}}$) in a coordinate-independent manner, we adopt the impact parameter $b$ as a natural measure. This approach directly relates the deviation of $b$ from its critical value $b_{\mathrm{c}}$ to the small difference $(\rho_0 - \rho_{\mathrm{m}})$. In particular, 
expanding $b$ in Eq.~\eqref{eq:batrho0} around $\rho_0=\rho_{\mathrm{m}}$ and using the circular-orbit condition~\eqref{eq:ccond2}, which eliminates the linear term, we obtain
\begin{align}
b=b_{\mathrm{c}} \left[\:\!
1-\frac{e^{2\:\!\gamma_{\mathrm{m}} }V''_{\mathrm{m}}}{4W_{\mathrm{m}}^2} (\rho_0-\rho_{\mathrm{m}})^2+O\left(
\Big(\frac{\rho_0}{\rho_{\mathrm{m}}}-1\Big)^3
\right)
\:\!\right].
\label{eq:bseries}
\end{align}
Now, we assume that $R'_{\mathrm{m}}\neq 0$ in what follows (see Appendix~\ref{sec:A} for the case $R'_{\mathrm{m}}= 0$). Similarly, the coefficients $c_1$ and $c_2$ can be expanded in terms of $(\rho_0-\rho_{\mathrm{m}})$ as follows:
\begin{align}
c_1&=-\frac{R'_{\mathrm{m}} V''_{\mathrm{m}}}{R_{\mathrm{m}}} (\rho_0-\rho_{\mathrm{m}})
+O\left(
\Big(\frac{\rho_0}{\rho_{\mathrm{m}}}-1\Big)^2
\right),
\label{eq:c1A}
\\
c_2&=-\frac{V''_{\mathrm{m}}}{2}+O\left(
\frac{\rho_0}{\rho_{\mathrm{m}}}-1
\right).
\label{eq:c1B}
\end{align}
Alternatively, by inverting Eq.~\eqref{eq:bseries}, we can express Eqs.~\eqref{eq:c1A} and \eqref{eq:c1B} as
\begin{align}
c_1&=2 \:\!e^{\psi_{\mathrm{m}}-\gamma_{\mathrm{m}}} R'_{\mathrm{m}}\sqrt{-V''_{\mathrm{m}}}
\left(
\frac{b}{b_{\mathrm{c}}}-1
\right)^{1/2}+O\left( 
\frac{b}{b_{\mathrm{c}}}-1
\right),
\label{eq:c1b}
\\
c_2&=-\frac{V''_{\mathrm{m}}}{2}+O\left( 
\Big(\frac{b}{b_{\mathrm{c}}}-1\Big)^{1/2}
\right).
\label{eq:c2b}
\end{align}
Using Eqs.~\eqref{eq:c1b} and \eqref{eq:c2b}, we can expand Eq.~\eqref{eq:IDc} in terms of $\left(b/b_{\mathrm{c}}-1\right)$ as 
\begin{align}
I_{\mathrm{D}}(\rho_0)=
&-\sqrt{-\frac{2}{V''_{\mathrm{m}}}} \log \left(
\frac{b}{b_{\mathrm{c}}}-1
\right)
\cr
&+\sqrt{-\frac{2}{V''_{\mathrm{m}}}} \log \left(
-\frac{e^{2(\gamma_{\mathrm{m}}-\psi_{\mathrm{m}})}}{(R'_{\mathrm{m}})^2}V''_{\mathrm{m}}
\right)
+O\left(\left(
\frac{b}{b_{\mathrm{c}}}-1
\right)^{1/2}
\log \left(
\frac{b}{b_{\mathrm{c}}}-1
\right)\right). 
\end{align}
This result shows that the leading divergence is logarithmic. Consequently, the deflection angle in the SDL is given by
\begin{align}
\alpha(\rho_0)=-\bar{a} \log \left(
\frac{b}{b_{\mathrm{c}}}-1
\right)+\bar{b}+O\left(\left(
\frac{b}{b_{\mathrm{c}}}-1
\right)^{1/2}
\log \left(
\frac{b}{b_{\mathrm{c}}}-1
\right)\right),
\label{eq:SDL}
\end{align}
where the SDL coefficients $\bar{a}$ and $\bar{b}$ are defined as
\begin{align}
\bar{a}&=\sqrt{-\frac{2}{V''_{\mathrm{m}}}},
\label{eq:abar}
\\
\bar{b}&
=-\bar{a} \log \left[\:\!
\frac{\bar{a}^2}{2 } (R'_{\mathrm{m}})^2\:\!e^{2(\psi_{\mathrm{m}}-\gamma_{\mathrm{m}})}
\:\!\right]+I_{\mathrm{R}}(\rho_{\mathrm{m}})-\pi.
\label{eq:bbar}
\end{align}
Here, $\bar{a}$ quantifies the strength of the logarithmic divergence, while $\bar{b}$ represents the constant offset, i.e., the regular part of the deflection angle after subtracting the logarithmic divergence. These expressions are fundamental to determining the deflection angle in the SDL in a coordinate-invariant manner.

\section{Coordinate-invariant form of the strong deflection coefficients via local curvature}
\label{sec:4}
In this section, we formulate the deflection angle in the SDL~\eqref{eq:SDL}--\eqref{eq:bbar} using local and coordinate-invariant geometric quantities. To this end, we introduce the following tetrad:
\begin{align}
e_{(0)}^a&=e^{-\psi} \:\!(\partial/ \partial t)^a,
\label{eq:e0}
\\
e_{(1)}^a&=e^{\psi-\gamma}\:\!(\partial/\partial \rho)^a,
\\
e_{(2)}^a&=e^{\psi-\gamma}\:\!(\partial/\partial \zeta)^a,
\\
e_{(3)}^a&=\frac{e^{\psi}}{W}\:\!(\partial/\partial \varphi)^a.
\label{eq:e3}
\end{align}
Here and hereafter, indices with parentheses, such as $(\mu)$ and $(\nu)$ with $\mu, \nu=0, 1, 2, 3$, denote the components in the orthonormal tetrad basis, whereas unparenthesized indices represent the abstract tensor indices introduced in the Introduction.
Here, $e_{(0)}^a$ represents the four-velocity of static observers. The tetrad components of the Einstein tensor, $G_{(\mu)(\nu)}=G_{ab}e_{(\mu)}^a e_{(\nu)}^b$, have the following nonzero components:
\begin{align}
G_{(0)(0)}&=e^{2\:\!(\psi-\gamma)}\left(
2\:\!(\psi_{\rho\rho}+\psi_{\zeta\zeta})
-\psi_{\rho}^2-\psi_{\zeta}^2
-\gamma_{\rho\rho}-\gamma_{\zeta\zeta}
+\frac{2(\psi_{\rho} W_{\rho}+\psi_{\zeta} W_{\zeta})}{W}
-\frac{W_{\rho\rho}+W_{\zeta\zeta}}{W}
\right),
\label{eq:G00}
\\
G_{(1)(1)}&=e^{2\:\!(\psi-\gamma)}\left(
-\psi_{\rho}^2+\psi_{\zeta}^2
+\frac{\gamma_{\rho}W_{\rho}-\gamma_{\zeta} W_{\zeta}}{W}
+\frac{W_{\zeta\zeta}}{W}
\right),
\\
G_{(2)(2)}&=e^{2\:\!(\psi-\gamma)}\left(
\psi_{\rho}^2-\psi_{\zeta}^2
-\frac{\gamma_{\rho}W_{\rho}-\gamma_{\zeta}W_{\zeta}}{W}
+\frac{W_{\rho\rho}}{W}
\right),
\\
G_{(1)(2)}&=e^{2\:\!(\psi-\gamma)}\left(
-2\:\!\psi_{\rho} \psi_{\zeta}
+\frac{\gamma_{\rho} W_{\zeta}+\gamma_{\zeta} W_{\rho}}{W}
-\frac{W_{\rho\zeta}}{W}
\right),
\\
G_{(3)(3)}&=e^{2\:\!(\psi-\gamma)}\left(
\psi_{\rho}^2+\psi_{\zeta}^2+\gamma_{\rho\rho}+\gamma_{\zeta\zeta}
\right),
\end{align}
with subscripts indicating partial differentiation (e.g., $\psi_{\rho}\equiv \partial \psi/\partial \rho$ and $\psi_{\rho\rho}\equiv \partial^2 \psi/\partial \rho^2$).

Let $C_{abcd}$ denote the Weyl tensor, which encodes the free gravitational field in the spacetime. The electric part of the Weyl tensor with respect to the four-velocity $e_{(0)}^a$ (see, e.g., Ref.~\cite{Stephani:2003tm}) is defined by
\begin{align}
E_{ab}=C_{acbd}\:\!e_{(0)}^ce_{(0)}^d. 
\end{align}
The nonzero tetrad components of $E_{ab}$, defined by $E_{(\mu)(\nu)}=E_{ab}e_{(\mu)}^a e_{(\nu)}^b$, are given by 
\begin{align}
E_{(1)(1)}=\frac{e^{2\:\!(\psi-\gamma)}}{6}\bigg[\:\!&
4\:\!\psi_{\rho\rho}-2\:\!\psi_{\zeta\zeta}
+8\:\!\psi_{\rho}^2-4\:\!\psi_{\zeta}^2
-\gamma_{\rho\rho}-\gamma_{\zeta\zeta}
-6\left(\psi_{\rho} \gamma_{\rho}-\psi_{\zeta} \gamma_{\zeta}
\right)
\cr
&-\frac{W_\rho\left(2\:\!\psi_\rho-3\:\!\gamma_{\rho}\right)}{W}
-\frac{W_{\zeta}\left(2\:\!\psi_{\zeta}+3\:\!\gamma_{\zeta}\right)}{W}
-\frac{W_{\rho\rho}-2 W_{\zeta\zeta}}{W}
\:\!\bigg],
\\
E_{(2)(2)}=\frac{e^{2\:\!(\psi-\gamma)}}{6}\bigg[\:\!&
4\:\!\psi_{\zeta\zeta}-2\:\!\psi_{\rho\rho}
+8\:\!\psi_{\zeta}^2-4\:\!\psi_{\rho}^2
-\gamma_{\rho\rho}-\gamma_{\zeta\zeta}
+6\left(\psi_{\rho} \gamma_{\rho}-\psi_{\zeta} \gamma_{\zeta}\right)
\cr
&-\frac{W_\rho\left(2\:\!\psi_{\rho}+3\:\!\gamma_{\rho}\right)}{W}
-\frac{W_{\zeta}\left(2\:\!\psi_{\zeta}-3\:\!\gamma_{\zeta}\right)}{W} 
-\frac{W_{\zeta\zeta}-2\:\!W_{\rho\rho}}{W}
\:\!\bigg],
\\
E_{(3)(3)}=\frac{e^{2\:\!(\psi-\gamma)}}{6}\bigg[\:\!&
-2\left(\psi_{\rho\rho}+\psi_{\zeta\zeta}\right)
-4\left(\psi_{\rho}^2+\psi_{\zeta}^2\right)
+2\left(\gamma_{\rho\rho}+\gamma_{\zeta\zeta}\right)
\cr
&+\frac{4\left(\psi_{\rho}W_{\rho}+\psi_{\zeta}W_{\zeta}\right)}{W}
-\frac{W_{\rho\rho}+W_{\zeta\zeta}}{W}
\:\!\bigg],
\\
E_{(1)(2)}=E_{(2)(1)}=e^{2\:\!(\psi-\gamma)}\bigg[\:\!&
\psi_{\rho\:\!\zeta}+2 \psi_{\rho}\psi_{\zeta}-\psi_{\rho}\gamma_{\zeta}-\psi_{\zeta}\gamma_{\rho}
+\frac{W_{\rho} \gamma_{\zeta}+W_{\zeta}\gamma_{\rho}-W_{\rho\:\!\zeta}}{2W}
\:\!\bigg].
\label{eq:E12}
\end{align}
Since the spacetime is static, the magnetic part of $C_{abcd}$ identically vanishes. Moreover, because the Weyl tensor is trace-free, we have
\begin{align}
E_{(1)(1)} + E_{(2)(2)} + E_{(3)(3)} = 0.
\end{align}

Due to the $\mathbb{Z}_2$ symmetry about the $\zeta=0$ plane [see Eq.~\eqref{eq:Z2}], we obtain 
\begin{align}
\psi_{\zeta}(\rho,0)=0, \quad 
\gamma_{\zeta}(\rho,0)=0, \quad
W_{\zeta}(\rho,0)=0. 
\label{eq:Z2symc}
\end{align}
Furthermore, any term involving an odd number of $\zeta$-derivatives (e.g., $\psi_{\rho\zeta}$) vanishes on the plane. Consequently, the off-diagonal components of both the electric part of the Weyl tensor and the Einstein tensor vanish on the plane (i.e., $E_{(1)(2)}=E_{(2)(1)}=0$ and $G_{(1)(2)}=0$ at $\zeta=0$). 

Using the curvature quantities defined above, the coordinate derivatives in Eq.~\eqref{eq:V"m} can be rewritten in terms of the tetrad components of the curvature tensors given in Eqs.~\eqref{eq:G00}--\eqref{eq:E12}.
Applying the $\mathbb{Z}_2$ symmetry conditions~\eqref{eq:Z2symc} on the $\zeta=0$ plane and evaluating the result at $\rho=\rho_{\mathrm{m}}$ with the circular-orbit condition~\eqref{eq:ccond2}, we obtain%
\footnote{A brief outline of how Eqs.~(39)--(48) are combined to obtain Eq.~\eqref{eq:V"mEG}
is presented in Appendix~\ref{sec:B}.}
\begin{align}
V''_{\mathrm{m}}=-2R^2_{\mathrm{m}}\left[\:\!
E_{(2)(2)}^{\mathrm{m}}-E_{(1)(1)}^{\mathrm{m}}-\frac{G_{(0)(0)}^{\mathrm{m}}+G_{(3)(3)}^{\mathrm{m}}}{2}
\:\!\right],
\label{eq:V"mEG}
\end{align}
where the subscript $\mathrm{m}$ of $E_{(\mu)(\nu)}$ and $G_{(\mu)(\nu)}$ indicates evaluation at $\rho=\rho_{\mathrm{m}}$ and $\zeta=0$.
 For the 
circular photon orbit
to be unstable, we require that $V''_{\mathrm{m}}<0$; hence,
\begin{align}
E_{(2)(2)}^{\mathrm{m}}-E_{(1)(1)}^{\mathrm{m}}>\frac{G_{(0)(0)}^{\mathrm{m}}+G_{(3)(3)}^{\mathrm{m}}}{2}.
\label{eq:UPCO}
\end{align}
This inequality establishes a local, coordinate-invariant relation between the Weyl and Einstein tensors (through their tetrad components) at the unstable 
circular photon orbit
(see also Ref.~\cite{Koga:2019uqd}). In the following section, we reinterpret this inequality in terms of the relationship between the matter field quantities and the tidal forces.

Finally, the SDL coefficients given in Eqs.~\eqref{eq:abar} and \eqref{eq:bbar} can be expressed in terms of curvature quantities as 
\begin{align}
\bar{a}&=\frac{1}{R_{\mathrm{m}} \sqrt{E_{(2)(2)}^{\mathrm{m}}-E_{(1)(1)}^{\mathrm{m}}-\frac{1}{2}\:\!\big(G_{(0)(0)}^{\mathrm{m}}+G_{(3)(3)}^{\mathrm{m}}\big)}},
\label{eq:abarG}
\\
\bar{b}&
=-\bar{a} \log\left[\:\!
\frac{\bar{a}^2}{12} R_{\mathrm{m}}^2\left(
6E_{(3)(3)}^{\mathrm{m}}+G_{(0)(0)}^{\mathrm{m}}-G_{(3)(3)}^{\mathrm{m}}+2\:\!G_{(1)(1)}^{\mathrm{m}}+2\:\!G_{(2)(2)}^{\mathrm{m}}
\right)
\:\!\right]+I_{\mathrm{R}}(\rho_{\mathrm{m}})-\pi,
\label{eq:bbarG}
\end{align}
where $R'_{\mathrm{m}}\neq 0$. These expressions show that the contributions from $I_{\mathrm{D}}$ to the SDL coefficients depend solely on $R_{\mathrm{m}}$ and the local, coordinate-invariant curvature quantities, $E_{(i)(i)}^{\mathrm{m}}$ and $G_{(\mu)(\mu)}^{\mathrm{m}}$. A comparison with the static and spherically symmetric analysis~\cite{Igata:2025taz} reveals a key difference: whereas the electric part of the Weyl tensor is absent in that case, it is essential in the present analysis. This underscores the influence of the tidal effects in more general spacetimes. 

We note that, although the contribution to $\bar{b}$ arising from $I_{\mathrm{D}}$ is fully determined by local curvature quantities, the remaining part $I_{\mathrm{R}}$ may involve global information about the spacetime. In particular, the decomposition into divergent and regular parts is not unique, and depends on the choice of the variable $z$.

\section{Relation between strong deflection limit coefficients and matter field quantities}
\label{sec:5}
In this section, we relate the SDL coefficients to the matter field quantities. We define a tensor $T_{(\mu)(\nu)}$ via the relation
\begin{align}
G_{(\mu)(\nu)}=8\pi \:\!T_{(\mu)(\nu)},
\label{eq:E-eqs}
\end{align}
which, in the context of general relativity, represents the Einstein equations with $T_{(\mu)(\nu)}$ interpreted as the energy-momentum tensor. In alternative theories of gravity, however, $T_{(\mu)(\nu)}$ may be understood as a more general tensor that not only encodes the energy-momentum distribution of matter fields but also incorporates additional curvature contributions. Without assuming a fixed physical interpretation for $T_{(\mu)(\nu)}$, we establish a framework in which the SDL coefficients remain independent of any specific gravitational theory. For simplicity, we will henceforth refer to $T_{(\mu)(\nu)}$ as the matter field quantities.

Let $T_{(\mu)(\nu)}^{\mathrm{m}}$ denote the value of $T_{(\mu)(\nu)}$ evaluated at $\rho=\rho_{\mathrm{m}}$ on the $\zeta=0$ plane (i.e., at the unstable 
circular photon orbit).
Then, $V''_{\mathrm{m}}$ given in Eq.~\eqref{eq:V"mEG} can be written as 
\begin{align}
V''_{\mathrm{m}}=-2R^2_{\mathrm{m}}\left[\:\!
E_{(2)(2)}^{\mathrm{m}}-E_{(1)(1)}^{\mathrm{m}}-4\pi \left(T_{(0)(0)}^{\mathrm{m}}+T_{(3)(3)}^{\mathrm{m}}\right)
\:\!\right].
\label{eq:V"mET}
\end{align}
For $V$ to exhibit a local maximum at $\rho=\rho_{\mathrm{m}}$ on the $\zeta=0$ plane, we require that $V''_{\mathrm{m}}<0$; equivalently, 
\begin{align}
E_{(2)(2)}^{\mathrm{m}}-E_{(1)(1)}^{\mathrm{m}}>4\pi \left(T_{(0)(0)}^{\mathrm{m}}+ T_{(3)(3)}^{\mathrm{m}}\right),
\label{eq:UPCO2}
\end{align}
as stated in Eq.~\eqref{eq:UPCO}. In other words, the existence of an unstable 
circular photon orbit
requires that the difference between the tidal components, $E_{(2)(2)}^{\mathrm{m}}-E_{(1)(1)}^{\mathrm{m}}$, is sufficiently large compared to the net contribution from the local matter fields $T_{(0)(0)}^{\mathrm{m}}+ T_{(3)(3)}^{\mathrm{m}}$. When the net matter contribution is negligible (e.g., typically in vacuum of general relativity), the tidal effects alone determine the orbital instability as
\begin{align}
E_{(2)(2)}^{\mathrm{m}}-E_{(1)(1)}^{\mathrm{m}}>0.
\label{eq:UPCO22}
\end{align}
Conversely, if the matter contribution is too large relative to the tidal difference, the condition in Eq.~\eqref{eq:UPCO2} is violated, and the circular photon orbit
becomes stable rather than unstable.

Finally, the SDL coefficients $\bar{a}$ and $\bar{b}$ can be expressed in terms of local matter field quantities as
\begin{align}
\bar{a}&=\frac{1}{R_{\mathrm{m}} \sqrt{E_{(2)(2)}^{\mathrm{m}}-E_{(1)(1)}^{\mathrm{m}}-4\pi\:\!\big(T_{(0)(0)}^{\mathrm{m}}+ T_{(3)(3)}^{\mathrm{m}}\big)}},
\label{eq:abarET}
\\
\bar{b}&
=-\bar{a} \log\left[\:\!
\frac{\bar{a}^2}{6} R_{\mathrm{m}}^2\left[\:\!
3E_{(3)(3)}^{\mathrm{m}}+4\pi \big(T_{(0)(0)}^{\mathrm{m}}-T_{(3)(3)}^{\mathrm{m}}+2\:\!T_{(1)(1)}^{\mathrm{m}}+2\:\!T_{(2)(2)}^{\mathrm{m}}\big)
\:\!\right]
\:\!\right]+I_{\mathrm{R}}(\rho_{\mathrm{m}})-\pi,
\label{eq:bbarET}
\end{align}
where $R'_{\mathrm{m}}\neq 0$. These expressions reveal that the contributions from $I_{\mathrm{D}}$ to the SDL coefficients are determined locally by $R_{\mathrm{m}}$, $E_{(i)(i)}^{\mathrm{m}}$, and the matter field quantities $T_{(\mu)(\mu)}^{\mathrm{m}}$. In particular, the coefficient $\bar{a}$ is entirely governed by the balance between the tidal contribution, $R_{\mathrm{m}}^2(E_{(2)(2)}^{\mathrm{m}}-E_{(1)(1)}^{\mathrm{m}})$, and the matter field contribution, $4\pi R_{\mathrm{m}}^2(T_{(0)(0)}^{\mathrm{m}}+ T_{(3)(3)}^{\mathrm{m}})$. Similarly, the contribution to $\bar{b}$ associated with $I_{\mathrm{D}}$ depends on $\bar{a}$ and the local balance between the tidal effects, $3R_{\mathrm{m}}^2E_{(3)(3)}^{\mathrm{m}}$, and the matter effects, $4\pi R_{\mathrm{m}}^2(T_{(0)(0)}^{\mathrm{m}}-T_{(3)(3)}^{\mathrm{m}}+2\:\!T_{(1)(1)}^{\mathrm{m}}+2\:\!T_{(2)(2)}^{\mathrm{m}})$.

The coordinate-invariant expressions for the SDL coefficients derived above 
indicate a direct link between the curvature and matter quantities
and observable lensing features near compact objects. If $R_{\mathrm{m}}$ is known, the observationally inferred $\bar{a}$ directly reflects information about the local tidal structure and matter distribution. In combination with accurate modeling of the background geometry, even the subleading term $\bar{b}$ may provide additional insights into the curvature and matter profile near the unstable 
circular photon orbit.

We focus on the special case where the following relations are satisfied:
\begin{align}
T_{(1)(1)}^{\mathrm{m}}+T_{(2)(2)}^{\mathrm{m}}=T_{(3)(3)}^{\mathrm{m}}=-T_{(0)(0)}^{\mathrm{m}},
\label{eq:mbala}
\end{align}
which are trivially satisfied in the vacuum of general relativity. Under these conditions, the expressions~\eqref{eq:abarET} and \eqref{eq:bbarET} reduce to 
\begin{align}
\bar{a}&=\frac{1}{R_{\mathrm{m}} \sqrt{E_{(2)(2)}^{\mathrm{m}}-E_{(1)(1)}^{\mathrm{m}}}},
\label{eq:abarETvac}
\\
\bar{b}&=-\bar{a} \log\left(
\frac{\bar{a}^2}{2} R_{\mathrm{m}}^2
E_{(3)(3)}^{\mathrm{m}}
\right)+I_{\mathrm{R}}(\rho_{\mathrm{m}})-\pi.
\label{eq:bbarETvac}
\end{align}
These results indicate that, when the matter field quantities satisfy the balance condition given in Eq.~\eqref{eq:mbala}, the SDL coefficients are determined solely by the free gravitational field.

\section{Strong deflection limit coefficients and Newman--Penrose scalars}
\label{sec:6}
To clarify the connection between the SDL coefficients and the intrinsic curvature of spacetime, we recast these coefficients in terms of NP scalars. This coordinate-invariant formulation directly links the observable features of gravitational lensing in the strong field limit to the fundamental curvature components of spacetime.

We introduce a null tetrad constructed from the orthonormal basis given in Eqs.~\eqref{eq:e0}--\eqref{eq:e3}. Specifically, we define
\begin{align}
l^a&=\frac{1}{\sqrt{2}}\Bigl(e_{(0)}^a+e_{(3)}^a\Bigr),
\label{eq:l^a}
\\ 
n^a&=\frac{1}{\sqrt{2}}\Bigl(e_{(0)}^a-e_{(3)}^a\Bigr),
\\
m^a&=\frac{1}{\sqrt{2}}\Bigl(e_{(1)}^a+i \:\!e_{(2)}^a\Bigr),
\\ 
\bar{m}^a&=\frac{1}{\sqrt{2}}\Bigl(e_{(1)}^a-i \:\!e_{(2)}^a\Bigr),
\label{eq:mb^a}
\end{align}
where $i$ denotes the imaginary unit. These null vectors satisfy the standard normalization conditions $l^a n_a=-1$ and $m^a \bar{m}_a=1$, with all other scalar products vanishing. 

We now project curvature tensors onto this null tetrad to obtain the NP scalars. Following Ref.~\cite{Stephani:2003tm}, we define the Weyl scalars as
\begin{align}
\Psi_0&=C_{abcd}\:\! l^a \:\!m^b\:\!l^c\:\!m^d,
\\
\Psi_1&=C_{abcd}\:\!l^a\:\!n^b\:\!l^c\:\!m^d,
\\
\Psi_2&=C_{abcd}\:\!l^a\:\! m^b \:\!\bar{m}^c\:\! n^d,
\\
\Psi_3&=C_{abcd}\:\!l^a \:\!n^b\:\!\bar{m}^c\:\!n^d,
\\
\Psi_4&=C_{abcd}\:\!\bar{m}^a\:\!n^b\:\!\bar{m}^c\:\!n^d.
\end{align}
These scalars encode the free gravitational field in a coordinate-invariant manner. 
The relation between the electric part of the Weyl tensor $E_{(i)(j)}$ and the NP Weyl scalars $\Psi_A$ ($A=0,1,\ldots, 4$) follows from the standard decomposition of the Weyl tensor in the NP formalism (see Ref.~\cite{Stephani:2003tm}). In a static spacetime with $\mathbb{Z}_2$ symmetry, the conditions $\Psi_1=\Psi_3=0$ and $\Psi_0=\Psi_4$ (real on the $\zeta=0$ plane) hold, and the nonvanishing components of $E_{(i)(j)}$ on the $\zeta=0$ plane are given by
\begin{align}
E_{(i)(j)}=\left[\:\!
\begin{array}{ccc}
\Psi_4-\Psi_2&0&0\\
0&-\Psi_4-\Psi_2&0\\
0&0&2\Psi_2
\end{array}
\:\!\right].
\end{align}

We also introduce the NP-Ricci scalars as
\begin{align}
\Phi_{00}&=\frac{1}{2}R_{ab}\:\!l^a l^b,
\\
\Phi_{11}&=\frac{1}{4} R_{ab} (l^a n^b+m^a \bar{m}^b),
\\
\Phi_{22}&=\frac{1}{2} R_{ab}n^a n^b,
\\
\Phi_{01}&=\frac{1}{2} R_{ab}\:\!l^a m^b,
\\
\Phi_{02}&=\frac{1}{2} R_{ab}\:\! m^a m^b,
\\
\Phi_{12}&=\frac{1}{2} R_{ab}\:\!m^a n^b.
\end{align}
These scalars encode the local matter contributions via the Einstein equations. 
By using $G_{ab}=R_{ab}-\frac{1}{2}R g_{ab}$ and the null tetrad conditions, the Ricci scalar $R$ drops out of $\Phi_{AB}$, so that $R_{ab}$ in $\Phi_{AB}$ can be simply replaced with $G_{ab}$. For a static spacetime, we find that $\Phi_{01}=0$, $\Phi_{12}=0$, and $\Phi_{00}=\Phi_{22}$. Moreover, $\mathbb{Z}_2$ symmetry about the $\zeta=0$ plane enforces that $\Phi_{02}$ is real on that plane. Under these conditions with Eqs.~\eqref{eq:l^a}--\eqref{eq:mb^a}, the nonvanishing NP-Ricci scalars on the $\zeta=0$ plane are related to the tetrad components of the Einstein tensor $G_{(\mu)(\nu)}$ as
\begin{align}
\Phi_{00}&=\Phi_{22}=\frac{G_{(0)(0)}+G_{(3)(3)}}{4}, 
\\ 
\Phi_{11}&=\frac{G_{(0)(0)}-G_{(3)(3)}+G_{(1)(1)}+G_{(2)(2)}}{8},
\\
\Phi_{02}&=\frac{G_{(1)(1)}-G_{(2)(2)}}{4}.
\end{align}

We denote by $\Psi_{0}^{\mathrm{m}}$, $\Psi_{2}^{\mathrm{m}}$, and $\Psi_{4}^{\mathrm{m}}$ the Weyl scalars evaluated at $\rho=\rho_{\mathrm{m}}$ on the $\zeta=0$ plane (i.e., at the unstable 
circular photon orbit).
Similarly, the NP-Ricci scalars at the same location are denoted by $\Phi_{00}^{\mathrm{m}}$, $\Phi_{11}^{\mathrm{m}}$, and $\Phi_{22}^{\mathrm{m}}$. Recalling Eqs.~\eqref{eq:V"mEG}, we find that $V''_{\mathrm{m}}$ is given by 
\begin{align}
V''_{\mathrm{m}}=4 R_{\mathrm{m}}^2 \left(
\Psi_{0}^{\mathrm{m}}+\Phi_{00}^{\mathrm{m}}
\right). 
\end{align}
Since the orbit is unstable [i.e., $V''_{\mathrm{m}}<0$ as required in Eq.~\eqref{eq:UPCO}], it follows that 
\begin{align}
\Psi_{0}^{\mathrm{m}}+\Phi_{00}^{\mathrm{m}}<0,
\end{align}
which encapsulates the necessary balance between the free gravitational field---encoded in $\Psi_0^{\mathrm{m}}$ (or $\Psi_4^{\mathrm{m}}$)---and the local matter contributions---captured by $\Phi_{00}^{\mathrm{m}}$. Thus, the NP formalism greatly simplifies the criterion for the instability of 
circular photon orbits.

Finally, we obtain the alternative forms of the SDL coefficients in terms of the NP scalars as
\begin{align}
\bar{a}&=\frac{1}{R_{\mathrm{m}} \sqrt{-2\big(
\Psi_0^{\mathrm{m}}+\Phi_{00}^{\mathrm{m}}
\big)}},
\label{eq:abarNP}
\\
\bar{b}&=-\bar{a}\log \left[\:\!
\bar{a}^2 R_{\mathrm{m}}^2\left(\Psi_2^{\mathrm{m}}+\Phi_{11}^{\mathrm{m}}-\frac{\mathcal{R}^{\mathrm{m}}}{24}\right)
\:\!\right]+I_{\mathrm{R}}(\rho_{\mathrm{m}})-\pi.
\label{eq:bbarNP}
\end{align}
where $\mathcal{R}^{\mathrm{m}}$ denotes the Ricci scalar at $\rho=\rho_{\mathrm{m}}$ and $\zeta=0$. 
The coefficient $\bar{a}$ scales only with $R_{\mathrm{m}}^2(\Psi_0^{\mathrm{m}}+\Phi_{00}^{\mathrm{m}})$, revealing that the tidal effects are captured by $\Psi_0^{\mathrm{m}}$, while the matter contributions are reflected in $\Phi_{00}^{\mathrm{m}}$. Similarly, the contribution to $\bar{b}$ from $I_{\mathrm{D}}$ depends on $\bar{a}$ and on $R_{\mathrm{m}}^2(\Psi_2^{\mathrm{m}}+\Phi_{11}^{\mathrm{m}}-\mathcal{R}^{\mathrm{m}}/24)$. These expressions relate the observable deflection angle in the SDL to specific combinations of NP scalars in the strong gravitational field regime. In particular, if the circumferential radius $R_{\mathrm{m}}$ is known, the combination $\Psi_0^{\mathrm{m}}+\Phi_{00}^{\mathrm{m}}$ can be extracted from observational data. Furthermore, by adopting a specific model, we can estimate $\Psi_2^{\mathrm{m}}+\Phi_{11}^{\mathrm{m}}-\mathcal{R}^{\mathrm{m}}/24$ based on observations. This NP-based formulation thus provides a coordinate-invariant framework for connecting lensing observations with local spacetime curvature, and may serve as a practical tool for probing strong gravity regions near compact objects.

\section{Evaluation of the formalism in specific geometries}
\label{sec:7}
We illustrate the consistency and physical relevance of the general formalism by evaluating the SDL coefficients in several explicit 
static and axisymmetric
spacetimes, showing how local curvature and matter fields determine light propagation near unstable photon orbits.

\subsection{Zipoy--Voorhees spacetimes}
\label{sec:7A}
We consider the Zipoy--Voorhees spacetimes, which satisfy the vacuum Einstein equations. The metric functions in Eq.~\eqref{eq:metric} are given by
\begin{align}
e^{2\psi}&=\left(
\frac{R_{+}+R_{-}-2\ell}{R_{+}+R_{-}+2\ell}
\right)^{\delta},
\quad
e^{2\gamma}=\left(
\frac{(R_{+}+R_{-})^2-4\ell^2}{4 R_{+} R_{-}}
\right)^{\delta^2},
\quad
W=\rho,
\end{align}
with
\begin{align}
R_\pm=\sqrt{\rho^2+(\zeta\pm \ell)^2}.
\end{align}
Here, $\ell$ and $m$ are positive constants, and the deformation parameter $\delta$ is defined by
\begin{align}
\delta=\frac{m}{\ell}.
\end{align}
When $\delta=1$, this metric 
reduces
to the Schwarzschild metric.

Solving Eq.~\eqref{eq:ccond2} yields the coordinate radius of the 
circular photon orbit
as
\begin{align}
\rho_{\mathrm{m}}=m\sqrt{4-\frac{1}{\delta^2}}, 
\end{align}
which implies that $\rho_{\mathrm{m}}$ exists only for $\delta>1/2$. 
The corresponding circumferential radius of the 
circular photon orbit, 
$R_{\mathrm{m}}$, and the critical impact parameter, given in Eq.~\eqref{eq:bc}, are
\begin{align}
R_{\mathrm{m}}
&=\left(2+\frac{1}{\delta}\right)\left(\frac{2\delta-1}{2\delta+1}\right)^{\frac{1-\delta}{2}}m,
\label{eq:RmZV}
\\
b_{\mathrm{c}}&=\left(2+\frac{1}{\delta}\right)\left(
\frac{2\delta-1}{2\delta+1}
\right)^{\frac{1-2\delta}{2}}m. 
\label{eq:bcZV}
\end{align}
Note that these results satisfy the following relation:
\begin{align}
b_{\mathrm{c}}=\left(\frac{2\delta+1}{2\delta-1}\right)^{\frac{\delta}{2}}R_{\mathrm{m}}.
\end{align}
When $\delta=1$ (i.e., the Schwarzschild case), $R_{\mathrm{m}}=3m$ and $b_{\mathrm{c}}=3\sqrt{3}m$. The second derivative of the effective potential at the 
circular photon orbit 
is given by
\begin{align}
V''_{\mathrm{m}}=-2 R_{\mathrm{m}}^2 (E_{(2)(2)}^{\mathrm{m}}-E_{(1)(1)}^{\mathrm{m}})=-2 \left(\frac{4\delta^2-1}{4\delta^2}\right)^{1-\delta^2}<0,
\end{align}
which indicates that the 
circular photon orbit
is unstable. 

The tetrad components of the Einstein tensor identically vanish. The nontrivial components of $E_{(\mu)(\nu)}^{\mathrm{m}}$ are
\begin{align}
R_{\mathrm{m}}^2E_{(1)(1)}^{\mathrm{m}}&=-\frac{5\delta^2-1}{8\delta^2}\left(\frac{4\delta^2}{4\delta^2-1}\right)^{\!\delta^2},
\\
R_{\mathrm{m}}^2E_{(2)(2)}^{\mathrm{m}}&=\frac{3\delta^2-1}{8\delta^2}\left(\frac{4\delta^2}{4\delta^2-1}\right)^{\!\delta^2},
\\
R_{\mathrm{m}}^2 E_{(3)(3)}^{\mathrm{m}}&=\frac{1}{4}\left(\frac{4\delta^2}{4\delta^2-1}\right)^{\delta^2}.
\end{align}
Correspondingly, the nontrivial NP scalars are given by
\begin{align}
R_{\mathrm{m}}^2 \Psi_0^{\mathrm{m}}&=R_{\mathrm{m}}^2 \Psi_4^{\mathrm{m}}=-\frac{1}{2}\left(\frac{4\delta^2-1}{4\delta^2}\right)^{1-\delta^2},
\label{eq:Psi4ZV}
\\
R_{\mathrm{m}}^2 \Psi_2^{\mathrm{m}}&=\frac{1}{8}\left(\frac{4\delta^2}{4\delta^2-1}\right)^{\delta^2}.
\end{align}

Finally, we obtain the explicit forms of Eqs.~\eqref{eq:abarETvac} and \eqref{eq:bbarETvac} as
\begin{align}
\bar{a}&=\left(\frac{4\delta^2}{4\delta^2-1}\right)^{\frac{1-\delta^2}{2}},
\\
\bar{b}&=\bar{a}\log \left(2\:\!\frac{4\delta^2-1}{\delta^2}\right)+I_{\mathrm{R}}(\rho_{\mathrm{m}})-\pi.
\end{align}
The result of $\bar{a}$ coincides with that of Ref.~\cite{Chakrabarty:2022fbd}, providing a concrete example that supports our main results. As mentioned in Sec.~\ref{sec:4}, the contribution from $I_{\mathrm{D}}$ to $\bar{b}$ depends on the choice of $z$.

In the static and spherically
symmetric limit, we obtain $\bar{a}=1$ and $\bar{b}=\log6+I_{\mathrm{R}}(\rho_{\mathrm{m}})-\pi$ (see, e.g., Refs.~\cite{Bozza:2002zj,Tsukamoto:2016jzh,Igata:2025taz}). This limiting case implies 
\begin{align}
E_{(2)(2)}^{\mathrm{m}}=
E_{(3)(3)}^{\mathrm{m}}=-\frac{
E_{(1)(1)}^{\mathrm{m}}}{2}=\frac{1}{3R_{\mathrm{m}}^2}.
\end{align}
Here, the equality $E_{(2)(2)}=E_{(3)(3)}$ follows directly from spherical symmetry, and the trace-free property of the Weyl tensor then yields $E_{(1)(1)}=-2E_{(2)(2)}$. This degeneracy in $E_{(i)(j)}$, together with the absence of matter fields, results in the universal value $\bar{a}=1$. Under this degeneracy, the NP scalars reduce to
\begin{align}
R_{\mathrm{m}}^2 \Psi_4^{\mathrm{m}}=R_{\mathrm{m}}^2 \Psi_0^{\mathrm{m}}=-\frac{1}{2},
\quad
R_{\mathrm{m}}^2 \Psi_2^{\mathrm{m}}=\frac{1}{6}.
\end{align}

\subsection{Reissner--Nordstr\"om spacetimes}
\label{sec:7B}
We consider the Reissner--Nordstr\"om spacetimes characterized by mass $M$ and electric charge $Q$. In the metric form presented in Eq.~\eqref{eq:metric}, the metric functions are given by (see, e.g., Ref.~\cite{Griffiths:2012})
\begin{align}
e^{2\:\!\psi}=\frac{(R_{+}+R_{-})^2-4 d^2}{(R_{+}+R_{-}+2M)^2}, \quad
e^{2\:\!\gamma}=\frac{(R_{+}+R_{-})^2-4 d^2}{4R_{+}R_{-}}, \quad
W=\rho,
\end{align}
where $d=\sqrt{M^2-Q^2}$ and 
$R_\pm=\sqrt{\rho^2+(\zeta\pm d)^2}$.
When $Q=0$, this metric reduces to the Schwarzschild metric. 

Solving Eq.~\eqref{eq:ccond2} yields the coordinate radius of
a circular photon orbit:
\begin{align}
\rho_{\mathrm{m}}=\sqrt{(R_{\mathrm{m}}-M)^2-d^2},
\end{align}
where $R_{\mathrm{m}}$ is given by
\begin{align}
R_{\mathrm{m}}=\frac{3M+\sqrt{9M^2-8Q^2}}{2}, 
\label{eq:RmRN}
\end{align}
indicating that 
the circular photon orbit exists
only when $0\le Q^2 \le 9M^2/8$. 
The critical impact parameter \eqref{eq:bc} is 
\begin{align}
b_{\mathrm{c}}=R_{\mathrm{m}}\sqrt{\frac{3}{1-Q^2/R_{\mathrm{m}}^2}}. 
\label{eq:bcRN}
\end{align}
The second derivative of the effective potential at the 
circular photon orbit
is given by
\begin{align}
V''_{\mathrm{m}}
=-2\left(2-\frac{3M}{R_{\mathrm{m}}}\right).
\end{align}
The condition $V''_{\mathrm{m}}<0$ further restricts the electric charge to the range $0\le Q^2 < 9M^2/8$.
This confirms that $R_{\mathrm{m}}$ corresponds to the unstable circular photon orbit. Note that for $M^2< Q^2\le 9M^2/8$ (i.e., the over-extremal regimes), another circular photon orbit appears at $R_{\mathrm{m},-}=(3M-\sqrt{9M^2-8Q^2})/2$, which corresponds to a stable circular photon orbit. In the extremal case $Q^2=M^2$, this orbit coincides with the horizon radius and therefore does not represent an independent stable circular photon orbit outside the horizon.

The nontrivial components of $T_{(\mu)(\nu)}^{\mathrm{m}}$ are given by
\begin{align}
T_{(0)(0)}^{\mathrm{m}}=T_{(2)(2)}^{\mathrm{m}}=T_{(3)(3)}^{\mathrm{m}}=-T_{(1)(1)}^{\mathrm{m}}=\frac{3M-R_{\mathrm{m}}}{16\pi R_{\mathrm{m}}^3}.
\end{align}
From these expressions, the corresponding components of $G_{(\mu)(\mu)}^{\mathrm{m}}$ can be derived. The nontrivial components of $E_{(\mu)(\nu)}^{\mathrm{m}}$ are given by
\begin{align}
E_{(2)(2)}^{\mathrm{m}}
=E_{(3)(3)}^{\mathrm{m}}
=-\frac{E_{(1)(1)}^{\mathrm{m}}}{2}
=\frac{R_{\mathrm{m}}-M}{2R_{\mathrm{m}}^3}, 
\end{align}
which reflects a degeneracy of the Weyl tensor due to spherical symmetry. Correspondingly, the nontrivial NP scalars are given by
\begin{align}
\Psi_4^{\mathrm{m}}=\Psi_0^{\mathrm{m}}=-3\Psi_2^{\mathrm{m}}=-\frac{3(R_{\mathrm{m}}-M)}{4R_{\mathrm{m}}^3},
\label{eq:Psi4RN}
\end{align}
and
\begin{align}
\Phi_{00}^{\mathrm{m}}=\Phi_{22}^{\mathrm{m}}=\frac{Q^2}{2R_{\mathrm{m}}^4},
\label{eq:Phi00RN}
\quad
\Phi_{11}^{\mathrm{m}}=0, \quad \mathcal{R}^{\mathrm{m}}=0.
\end{align}

Finally, the coefficients $\bar{a}$ and $\bar{b}$ given by Eqs.~\eqref{eq:abarET} and \eqref{eq:bbarET}, or equivalently, Eqs.~\eqref{eq:abarNP} and \eqref{eq:bbarNP}, are
\begin{align}
\bar{a}&=\sqrt{\frac{R_{\mathrm{m}}}{2R_{\mathrm{m}}-3M}},
\\
\bar{b}&=\bar{a}\log \left(
8-\frac{4M}{R_{\mathrm{m}}-M}
\right)+I_{\mathrm{R}}(\rho_{\mathrm{m}})-\pi.
\end{align}
These results coincide with those obtained in Refs.~\cite{Eiroa:2002mk,Tsukamoto:2016oca,Igata:2025taz}, providing a consistency check for both $\bar{a}$ and $\bar{b}$.
When $Q=M$, the coefficients reduce to $\bar{a}=\sqrt{2}$ and $\bar{b}=\sqrt{2}\log 4+I_{\mathrm{R}}(\rho_{\mathrm{m}})-\pi$.

\subsection{Majumdar--Papapetrou dihole spacetimes}
\label{sec:7C}
We consider the Majumdar--Papapetrou dihole spacetimes, in which two extremal Reissner--Nordstr\"om black holes of equal mass $M$ are held in static equilibrium at a separation $a$. In the metric form presented in Eq.~\eqref{eq:metric}, the metric functions are given by
\begin{align}
e^{2\psi}=\left(
1+\frac{M}{R_+}+\frac{M}{R_-}
\right)^{-2}, \quad \gamma=0, \quad W=\rho,
\end{align}
where $R_\pm=\sqrt{\rho^2+(\zeta\pm a)^2}$. When $a=0$, the solution reduces to a single extremal Reissner--Nordstr\"om black hole with total mass $2M$ and charge $2M$. In what follows, we set $M=1$.

Solving Eq.~\eqref{eq:ccond2} yields three distinct solutions for the coordinate radius of the 
circular photon orbit
on the plane $\zeta=0$. Here, we select the branch corresponding to the unstable orbit, which is given by~\cite{Patil:2016oav,Shipley:2016omi,Assumpcao:2018bka}
\begin{align}
\rho_{\mathrm{m}}=\sqrt{\frac{4}{9}\left(1+2 \cos \left[\:\!
\frac{1}{3}\arccos\left(
1-\frac{27a^2}{4}
\right)
\:\!\right]\right)^2-a^2}.
\end{align}
The instability of this orbit is confirmed by 
\begin{align}
V''_{\mathrm{m}}=2+\frac{8\rho_{\mathrm{m}}^2\left[\:\!
2\rho_{\mathrm{m}}^2+(a^2-2\rho_{\mathrm{m}}^2)\big(2+\sqrt{\rho_{\mathrm{m}}^2+a^2}\big)
\:\!\right]}{\left(\rho_{\mathrm{m}}^2+a^2\right)^2\big(2+\sqrt{\rho_{\mathrm{m}}^2+a^2}\big)^2}<0,
\end{align}
for $0\le a<a_\infty\equiv 2\sqrt{6}/9$~\cite{Nakashi:2019mvs}. 
The corresponding circumferential radius and the critical impact parameter are given by
\begin{align}
R_{\mathrm{m}}&=\sqrt{\rho_{\mathrm{m}}}\left(1+\frac{2}{\sqrt{\rho_{\mathrm{m}}+a^2}}\right),
\label{eq:RmMP}
\\
b_{\mathrm{c}}&=\sqrt{\rho_{\mathrm{m}}}\left(1+\frac{2}{\sqrt{\rho_{\mathrm{m}}+a^2}}\right)^2.
\label{eq:bcMP}
\end{align}

The nontrivial components of $T_{(\mu)(\nu)}^{\mathrm{m}}$ are
\begin{align}
T_{(0)(0)}^{\mathrm{m}}=T_{(2)(2)}^{\mathrm{m}}=T_{(3)(3)}^{\mathrm{m}}=-T_{(1)(1)}^{\mathrm{m}}=\frac{\rho_{\mathrm{m}}^2}{\pi \left(\rho_{\mathrm{m}}^2+a^2\right)^2\big(2+\sqrt{\rho_{\mathrm{m}}^2+a^2}\big)^2}.
\end{align}
From these expressions, the components of $G_{(\mu)(\mu)}^{\mathrm{m}}$ can be derived. The nontrivial components of $E_{(\mu)(\nu)}^{\mathrm{m}}$ are
\begin{align}
E_{(1)(1)}^{\mathrm{m}}&=\frac{2\rho_{\mathrm{m}}^2\left[\:\!
2a^2+(a^2-2\rho_{\mathrm{m}}^2)\sqrt{\rho_{\mathrm{m}}^2+a^2}
\:\!\right]}{R_{\mathrm{m}}^2\left(\rho_{\mathrm{m}}^2+a^2\right)^2\big(2+\sqrt{\rho_{\mathrm{m}}^2+a^2}\big)^2},
\\
E_{(2)(2)}^{\mathrm{m}}&=\frac{2\rho_{\mathrm{m}}^2\left[\:\!
-4a^2+(\rho_{\mathrm{m}}^2-2a^2)\sqrt{\rho_{\mathrm{m}}^2+a^2}
\:\!\right]}{R_{\mathrm{m}}^2\left(\rho_{\mathrm{m}}^2+a^2\right)^2\big(2+\sqrt{\rho_{\mathrm{m}}^2+a^2}\big)^2},
\\
E_{(3)(3)}^{\mathrm{m}}&=\frac{2\rho_{\mathrm{m}}^2\left[\:\!
2a^2+(\rho_{\mathrm{m}}^2+a^2)^{3/2}\:\!\right]}{R_{\mathrm{m}}^2\left(\rho_{\mathrm{m}}^2+a^2\right)^2\big(2+\sqrt{\rho_{\mathrm{m}}^2+a^2}\big)^2}.
\end{align}
Correspondingly, the nontrivial NP scalars are given by
\begin{align}
\Psi_4^{\mathrm{m}}&=\Psi_0^{\mathrm{m}}=\frac{3\rho_{\mathrm{m}}^2\left[\:\!
a^4-\rho_{\mathrm{m}}^4+2a^2\sqrt{\rho_{\mathrm{m}}^2+a^2}
\:\!\right]}{R_{\mathrm{m}}^2(\rho_{\mathrm{m}}^2+a^2)^{5/2}\big(2+\sqrt{\rho_{\mathrm{m}}^2+a^2}\big)^2},
\label{eq:Psi4MP}
\\
\Psi_2^{\mathrm{m}}&=\frac{\rho_{\mathrm{m}}^{2}\left[\:\!
2a^2+(\rho_{\mathrm{m}}^2+a^2)^{3/2}
\:\!\right]
}{R_{\mathrm{m}}^2(\rho_{\mathrm{m}}^2+a^2)^{2} \big(2+\sqrt{\rho_{\mathrm{m}}^2+a^2}\big)^2},
\end{align}
and
\begin{align}
\Phi_{00}^{\mathrm{m}}=\Phi_{22}^{\mathrm{m}}=-\Phi_{02}^{\mathrm{m}}=\frac{2\rho_{\mathrm{m}}^{2}}{R_{\mathrm{m}}^2\left(\rho_{\mathrm{m}}^2+a^2\right)^{2}\big(2+\sqrt{\rho_{\mathrm{m}}^2+a^2}\big)^{2}}, \quad 
\Phi_{11}^{\mathrm{m}}=0, \quad \mathcal{R}^{\mathrm{m}}=0.
\label{eq:Phi00MP}
\end{align}

Finally, the coefficients $\bar{a}$ and $\bar{b}$ given by Eqs.~\eqref{eq:abarET} and \eqref{eq:bbarET}, or equivalently, Eqs.~\eqref{eq:abarNP} and \eqref{eq:bbarNP}, are
\begin{align}
\bar{a}&=\left[\:\!-1-\frac{4\rho_{\mathrm{m}}^2\left[\:\!
2\rho_{\mathrm{m}}^2+(a^2-2\rho_{\mathrm{m}}^2)\big(2+\sqrt{\rho_{\mathrm{m}}^2+a^2}\big)
\:\!\right]}{\left(\rho_{\mathrm{m}}^2+a^2\right)^2\big(2+\sqrt{\rho_{\mathrm{m}}^2+a^2}\big)^2}\:\!\right]^{-1/2},
\\
\bar{b}&=\bar{a} \log\left[\:\!
\frac{2\rho_{\mathrm{m}}^{2}\left(\rho_{\mathrm{m}}^2+a^2\right)
\big(2+\sqrt{\rho_{\mathrm{m}}^2+a^2}\big)^4}{\bar{a}^2 R_{\mathrm{m}}^2\big(2a^2+(\rho_{\mathrm{m}}^2+a^2)^{3/2}\big)^2}
\:\!\right]+I_{\mathrm{R}}(\rho_{\mathrm{m}})-\pi.
\end{align}
We find that $\bar{a}$ coincides with the result in Ref.~\cite{Patil:2016oav}, providing 
another concrete example that supports our main results. As mentioned in Sec.~\ref{sec:4}, the contribution from $I_{\mathrm{D}}$ to $\bar{b}$ depends on the choice of $z$. When $a=0$, these coefficients reduce to those for $Q=M$ in Sec.~\ref{sec:7B}.

\section{Strong deflection limit and quasinormal modes}
\label{sec:8}
A well-established connection exists between QNM frequencies and the instability of 
circular photon orbits 
in spherically symmetric spacetimes, particularly in the eikonal limit~\cite{Ferrari:1984zz,Cardoso:2008bp}. The real part of the QNM frequency corresponds to the angular frequency of the 
circular photon orbit
$\Omega_{\mathrm{c}}=1/b_{\mathrm{c}}$, while the imaginary part is determined by the Lyapunov exponent $\lambda_{\mathrm{L}}$ characterizing the instability of the orbit
\begin{align}
\omega_{\mathrm{QNM}}=\Omega_{\mathrm{c}} \:\! l-i\left(n+\frac{1}{2}\right)\lambda_{\mathrm{L}},
\end{align}
where $n$ is the overtone number 
and 
$l$ is the angular momentum of the perturbation~\cite{Cardoso:2008bp}. The Lyapunov exponent is given by 
\begin{align}
\lambda_{\mathrm{L}}=\frac{1}{b_{\mathrm{c}}}\sqrt{-\frac{V''_{\mathrm{m}}}{2}}.
\end{align}
Note that this expression is rewritten using the present notation, for consistency with our SDL formalism, rather than the original expression in Ref.~\cite{Cardoso:2008bp}. Several studies have further explored a potential correspondence between QNM parameters and the SDL coefficient that governs the logarithmic divergence rate~\cite{Stefanov:2010xz,Raffaelli:2014ola}
\begin{align}
\lambda_{\mathrm{L}}=\frac{1}{b_{\mathrm{c}}\bar{a}},
\label{eq:stefanov}
\end{align}
which indicates that the Lyapunov exponent appearing in the QNM frequency expression is proportional to the inverse of the SDL coefficient $\bar{a}$. Notably, a similar approach has also been employed on the reflection-symmetric plane of axisymmetric spacetimes, where QNM frequencies are computed using equatorial 
circular photon orbits.
These results suggest a deep geometric link between wave dynamics and lensing in strong gravity. 

Recently, an explicit expression for $\lambda_{\mathrm{L}}$ in spherically symmetric spacetimes has been derived in terms of local geometric and matter field quantities~\cite{Igata:2025taz}:
\begin{align}
\lambda_{\mathrm{L}}=\frac{\sqrt{1-8\pi R_{\mathrm{m}}^2 (\rho_{\mathrm{m}}+\Pi_{\mathrm{m}})}}{b_{\mathrm{c}}}.
\label{eq:Lyasph}
\end{align}
Here, $\rho_{\mathrm{m}}$ and $\Pi_{\mathrm{m}}$ denote energy density and tangential pressure evaluated at the photon sphere with the areal radius $R_{\mathrm{m}}$. This result reveals the relation between the damping rate of QNMs and the local matter field quantities. 

Our results in the present paper extend the formula \eqref{eq:Lyasph} to 
static and axisymmetric
spacetimes. Using Eqs.~\eqref{eq:abarET} or \eqref{eq:abarNP}, we recast $\lambda_{\mathrm{L}}$ in the form of Eq.~\eqref{eq:stefanov} as 
\begin{align}
\lambda_{\mathrm{L}}=\frac{R_{\mathrm{m}}}{b_{\mathrm{c}}} \sqrt{E_{(2)(2)}^{\mathrm{m}}-E_{(1)(1)}^{\mathrm{m}}-4\pi\:\!\big(T_{(0)(0)}^{\mathrm{m}}+ T_{(3)(3)}^{\mathrm{m}}\big)},
\end{align}
or equivalently, 
\begin{align}
\lambda_{\mathrm{L}}=\frac{R_{\mathrm{m}} \sqrt{-2\big(
\Psi_0^{\mathrm{m}}+\Phi_{00}^{\mathrm{m}}
\big)}}{b_{\mathrm{c}}}.
\label{eq:Lyapunov}
\end{align}
These expressions are determined by the coordinate-independent, local curvature, and matter field quantities evaluated at the unstable circular photon orbit.
This indicates that the damping rate of QNMs is directly linked to the local geometric and matter field properties at the radius of the unstable circular photon orbit.

In the following, we test our geometric expression for the Lyapunov exponent in several 
static and axisymmetric
spacetimes. For each case, we evaluate the NP scalars at the 
circular photon orbit
and confirm agreement with the result obtained from the effective potential. Explicit computations are given in Sec.~\ref{sec:7}.

In the case of the Zipoy--Voorhees spacetimes in Sec.~\ref{sec:7A}, where $\Phi_{00}^{\mathrm{m}}=0$, substituting Eqs.~\eqref{eq:RmZV}, \eqref{eq:bcZV}, and \eqref{eq:Psi4ZV} into the formula~\eqref{eq:Lyapunov}, we obtain
\begin{align}
\lambda_{\mathrm{L}}=\frac{1}{2m\delta}\left(
\frac{4\delta^2-1}{4\delta^2}
\right)\left(\frac{2\delta-1}{2\delta+1}\right)^{\delta}.
\end{align}
Expanding around $\delta=1$, we obtain
\begin{align}
\lambda_{\mathrm{L}}= \frac{1}{3\sqrt{3} m}\left[\:\!
1+\left(
2\log \frac{2}{3}\right)(\delta-1)+O((\delta-1)^2)
\:\!\right],
\end{align}
which agrees with the result obtained in Ref.~\cite{Allahyari:2018cmg}. The leading term $1/(3\sqrt{3}m)$ corresponds to the value for the Schwarzschild case. 

In the case of the Reissner--Nordstr\"om spacetime in Sec.~\ref{sec:7B}, substituting Eqs.~\eqref{eq:RmRN}, \eqref{eq:bcRN}, and \eqref{eq:Psi4RN} into the formula~\eqref{eq:Lyapunov}, we obtain
\begin{align}
\lambda_{\mathrm{L}}=\frac{1}{R_{\mathrm{m}}}\sqrt{\left(\frac{2}{3}-\frac{M}{R_{\mathrm{m}}}\right)\left(1-\frac{Q^2}{R_{\mathrm{m}}^2}\right)}, 
\end{align}
which agrees with the result obtained in Ref.~\cite{Pradhan:2012rkk}. When $Q=M$, the expression reduces to $\lambda_{\mathrm{L}}=1/(4\sqrt{2}M)$.

In the case of the Majumdar--Papapetrou dihole spacetimes in Sec.~\ref{sec:7C}, substituting Eqs.~\eqref{eq:RmMP}, \eqref{eq:bcMP}, \eqref{eq:Psi4MP}, and \eqref{eq:Phi00MP} into the formula~\eqref{eq:Lyapunov}, we obtain
\begin{align}
\lambda_{\mathrm{L}}=\frac{\sqrt{d_{\mathrm{m}}^4(4+4d_{\mathrm{m}}-d_{\mathrm{m}}^2)-4a^2d_{\mathrm{m}}(6+5d_{\mathrm{m}})+4a^4(4+3d_{\mathrm{m}})}}{(2+d_{\mathrm{m}})^3\sqrt{d_{\mathrm{m}}^2-a^2}},
\end{align}
which agrees with the result obtained in Ref.~\cite{Assumpcao:2018bka}. When $a=0$, the expression reduces to $\lambda_{\mathrm{L}}=1/(8\sqrt{2})$.

These consistent results support the validity of the geometric formula~\eqref{eq:Lyapunov} for the Lyapunov exponent in 
static and axisymmetric
spacetimes.

\section{Summary and discussion}
\label{sec:9}
We have studied the deflection angle of photons in the SDL in 
static and axisymmetric
spacetimes with reflection symmetry across the equatorial plane. Following the standard method of isolating the logarithmic divergence, we have introduced a new variable that enabled a coordinate-invariant formulation of the SDL. 
We have shown that the two SDL coefficients, which characterize the logarithmic divergence and its offset, 
are determined by the second derivative of the effective potential evaluated at the unstable photon circular orbit. We further have demonstrated that this second derivative can be expressed in terms of local curvature and matter field quantities, yielding a coordinate-invariant expression for these coefficients. This local curvature consists of two distinct contributions: the tidal effects associated with the free gravitational field, described by the electric part of the Weyl tensor, and the matter-induced curvature, encoded in the Einstein tensor, representing the influence of matter fields. We 
have also recast the SDL coefficients in terms of NP scalars, offering a geometric perspective aligned with the null structure of the spacetime. This formulation suggests a promising framework for interpreting observational data in terms of fundamental curvature quantities, particularly through coordinate-invariant combinations of NP Weyl and Ricci scalars. 
We have confirmed the consistency of our formulation by applying it to several known spacetimes.

In static and spherically symmetric spacetimes, the SDL coefficients can be expressed solely in terms of the areal radius of the photon sphere and the local energy density and pressure~\cite{Igata:2025taz}. In contrast, in 
static and axisymmetric
spacetimes, the deflection angle in the SDL is governed not only by local matter fields but also by the free gravitational field encoded in tidal distortions, leading to a more intricate geometric structure. Nevertheless, the leading logarithmic divergence of the deflection angle is still determined entirely by local quantities with clear geometric meaning at the unstable 
circular photon orbit.
This ensures that the formulation retains its coordinate-invariant character.

These results provide a geometrically transparent foundation for connecting theoretical predictions with observations. By expressing the SDL coefficients entirely in terms of coordinate-invariant local quantities, the present formalism establishes a clear link between observable lensing features and the underlying geometry and matter distribution near the unstable 
circular photon orbit.
This, in turn, opens up the possibility of inferring local geometric or matter-field properties---such as tidal forces, energy density, and pressure---from precise measurements of strong gravitational lensing near compact objects.
Although the deflection angle itself is not directly measurable, it determines the observable quantities such as the image positions, magnifications, and time delays through the lens equation (see, e.g., Refs.~\cite{Bozza:2002zj, Virbhadra:1999nm}).
Therefore, the local curvature expressions of the SDL coefficients obtained in this paper can be directly applied to interpret these measurable lensing features in a coordinate-invariant manner.

Through the connection between the QNM and SDL, we have expressed the QNM frequency in terms of local curvature and matter field quantities. This formulation opens up the possibility of probing the local matter and geometry near the unstable 
circular photon orbit
through gravitational wave observations. Furthermore, our findings may shed light on a universal upper bound on chaos in thermal quantum field theory~\cite{Maldacena:2015waa,Hashimoto:2016dfz}, which provides an inequality between the Lyapunov exponent and the surface gravity of the horizon, a relation whose generalization to the photon sphere has recently been discussed~\cite{Giataganas:2024hil,Gallo:2024wju}.

At present, it remains challenging to observationally compare the quantities related to the deflection angle in the SDL with the QNM frequencies because the typical targets of these observations are different: observations of black hole shadows are feasible only for supermassive black holes, whereas QNM frequencies are detected from mergers of stellar-mass binary black holes. However, if the gravitational-wave source itself acts as a lens, it would in principle be possible to observe both phenomena within a single system. Although such a self-lensing scenario is beyond the scope of this work, which considers only static and axisymmetric systems, it would provide a natural setting in which the geometrical and wave aspects of strong gravity could be observed simultaneously.

The present formalism provides a practical framework for extracting physical information---such as energy density, pressure, and tidal structure---from the deflection angle near ultracompact objects. Applied to specific models, it may offer valuable insights into the physics of strong gravity and serve as a probe of alternative theories of gravity. A natural extension of this work is to generalize the formalism to stationary and axisymmetric spacetimes; this development is already underway and will be presented in a separate publication.

\begin{acknowledgments}
The author gratefully acknowledges useful discussions with Yohsuke Takamori, Tomohiro Harada, and Akihito Katsumata. This work was supported in part by JSPS KAKENHI Grants No.~JP22K03611, No.~JP23KK0048, and No.~JP24H00183 and by Gakushuin University. 
\end{acknowledgments}

\appendix

\section*{DATA AVAILABILITY}
No data were created or analyzed in this study.

\section{Analysis of the $R'_{\mathrm{m}}=0$ case}
\label{sec:A}
We consider the case $R'_{\mathrm{m}}=0$, which, for example, is encountered in wormhole geometries. Since this condition implies that $W'_{\mathrm{m}}=W_{\mathrm{m}}\psi'_{\mathrm{m}}$, combining it with Eq.~\eqref{eq:ccond2} yields $W'_{\mathrm{m}}=0$. In this case, the coefficients given in Eqs.~\eqref{eq:prec1} and \eqref{eq:prec2} are expanded about $\rho_0=\rho_{\mathrm{m}}$ as 
\begin{align}
c_1&=-\frac{R''_{\mathrm{m}} V''_{\mathrm{m}}}{R_{\mathrm{m}}} (\rho_0-\rho_{\mathrm{m}})^2+O\left(\Big(\frac{\rho_0}{\rho_{\mathrm{m}}}-1\Big)^3\right),
\label{eq:c1Bzero}
\\
c_2&=-2 V''_{\mathrm{m}}+\left(
-\frac{3}{2}\frac{R'''_{\mathrm{m}} V''_{\mathrm{m}}}{R''_{\mathrm{m}}}-\frac{V'''_{\mathrm{m}}}{2}
\right)(\rho_{0}-\rho_{\mathrm{m}})+O\left(\Big(\frac{\rho_0}{\rho_{\mathrm{m}}}-1\Big)^2\right). 
\label{eq:c2Bzero}
\end{align}
Inverting Eq.~\eqref{eq:bseries} allows us to rewrite Eqs.~\eqref{eq:c1Bzero} and \eqref{eq:c2Bzero} in terms of $(b/b_{\mathrm{c}}-1)$ as
\begin{align}
c_1&=4 R_{\mathrm{m}}R''_{\mathrm{m}} e^{2(\psi_{\mathrm{m}}-\gamma_{\mathrm{m}})}\left(
\frac{b}{b_{\mathrm{c}}}-1
\right)+O\left(
\bigg(\frac{b}{b_{\mathrm{c}}}-1\bigg)^{3/2}
\right),
\label{eq:c1bWH}
\\
c_2&=-2V''_{\mathrm{m}}-\frac{e^{\psi_{\mathrm{m}}-\gamma_{\mathrm{m}}}R_{\mathrm{m}}}{\sqrt{-V''_{\mathrm{m}}}}\left(
\frac{3R'''_{\mathrm{m}}V''_{\mathrm{m}}}{R''_{\mathrm{m}}}+\frac{V'''_{\mathrm{m}}}{2}
\right)\left(
\frac{b}{b_{\mathrm{c}}}-1
\right)^{1/2}+O\left(
\frac{b}{b_{\mathrm{c}}}-1
\right).
\label{eq:c2bWH}
\end{align}
Using Eqs.~\eqref{eq:IDc}, \eqref{eq:c1bWH}, and \eqref{eq:c2bWH}, we obtain the deflection angle in the SDL~\eqref{eq:SDL}, with the corresponding SDL coefficients given by
\begin{align}
\bar{a}&=\sqrt{-\frac{2}{V''_{\mathrm{m}}}},
\label{eq:abarWH}
\\
\bar{b}&
=-\bar{a} \log \left[\:\!\frac{\bar{a}^2}{4 } R_{\mathrm{m}}R''_{\mathrm{m}}e^{2(\psi_{\mathrm{m}}-\gamma_{\mathrm{m}})}
\:\!\right]+I_{\mathrm{R}}(\rho_{\mathrm{m}})-\pi.
\label{eq:bbarWH}
\end{align}
Note that the coefficient $\bar{a}$ is identical to that obtained in the $R'_{\mathrm{m}}\neq 0$ case; consequently, Eqs.~\eqref{eq:abarG}, \eqref{eq:abarET}, and \eqref{eq:abarNP} remain unchanged. On the other hand, $\bar{b}$ is expressed in terms of the Einstein tensor as
\begin{align}
\bar{b}
=-\bar{a} \log\left\{\:\!
\frac{
\bar{a}^2}{16} R_{\mathrm{m}}^2\left[\:\!
2(E_{(2)(2)}^{\mathrm{m}}-E_{(1)(1)}^{\mathrm{m}})-G_{(0)(0)}^{\mathrm{m}}-G_{(3)(3)}^{\mathrm{m}}+2G_{(2)(2)}^{\mathrm{m}}
\:\!\right]
\:\!\right\}+I_{\mathrm{R}}(\rho_{\mathrm{m}})-\pi, 
\end{align}
in terms of matter field quantities as
\begin{align}
\bar{b}
=-\bar{a} \log\left\{\:\!
\frac{
\bar{a}^2}{8} R_{\mathrm{m}}^2\left[\:\!
E_{(2)(2)}^{\mathrm{m}}-E_{(1)(1)}^{\mathrm{m}}-4\pi (T_{(0)(0)}^{\mathrm{m}}+T_{(3)(3)}^{\mathrm{m}}-2T_{(2)(2)}^{\mathrm{m}})
\:\!\right]
\:\!\right\}+I_{\mathrm{R}}(\rho_{\mathrm{m}})-\pi,
\end{align}
and in terms of the NP scalars as
\begin{align}
\bar{b}=-\bar{a}\log \left\{\:\!
\frac{\bar{a}^2}{8} R_{\mathrm{m}}^2\left[\:\!
-2(\Psi_0^{\mathrm{m}}+\Phi_{00}^{\mathrm{m}}-\Phi_{11}^{\mathrm{m}}+\Phi_{02}^{\mathrm{m}})-\frac{\mathcal{R}^{\mathrm{m}}}{4}
\:\!\right]
\:\!\right\}+I_{\mathrm{R}}(\rho_{\mathrm{m}})-\pi.
\end{align}

\section{Outline of the derivation of $V''_{\mathrm{m}}$ in terms of curvature quantities}
\label{sec:B}
In this appendix, we provide a brief outline of how the coordinate derivatives 
in Eq.~\eqref{eq:V"m} are reorganized into the curvature quantities appearing in 
Eq.~\eqref{eq:V"mEG}, using the tetrad components of the Einstein and Weyl tensors. 
Using the $\mathbb{Z}_2$ symmetry conditions~\eqref{eq:Z2symc} on the 
$\zeta = 0$ plane, the tetrad components of the Einstein tensor and the Weyl tensor simplify 
considerably. In particular, the combination 
$G_{(0)(0)}^{\mathrm m} + G_{(3)(3)}^{\mathrm m}$ takes the form
\begin{align}
G_{(0)(0)}^{\mathrm{m}}+G_{(3)(3)}^{\mathrm{m}}
=2e^{2\:\!(\psi_{\mathrm{m}}-\gamma_{\mathrm{m}})}\left[\:\!
\psi_{\rho\rho}^{\mathrm{m}}
+\psi_{\zeta\zeta}^{\mathrm{m}}
+\frac{(W_\rho^{\mathrm{m}})^2}{2W_{\mathrm{m}}^2}
-\frac{W_{\rho\rho}^{\mathrm{m}}+W_{\zeta\zeta}^{\mathrm{m}}}{2W_{\mathrm{m}}}
\:\!\right],
\end{align}
where we have used Eq.~\eqref{eq:ccond2} to eliminate $\psi_\rho^{\mathrm{m}}$. 
Similarly, the difference of the tetrad components of the electric part of 
the Weyl tensor on the $\zeta=0$ plane reduces to the following expression:
\begin{align}
E_{(2)(2)}^{\mathrm{m}}-E_{(1)(1)}^{\mathrm{m}}
=e^{2\:\!(\psi_{\mathrm{m}}-\gamma_{\mathrm{m}})}\left[\:\!
\psi_{\zeta\zeta}^{\mathrm{m}}-\psi_{\rho\rho}^{\mathrm{m}}
-\frac{(W_\rho^{\mathrm{m}})^2}{2W_{\mathrm{m}}^2}
+\frac{W_{\rho\rho}^{\mathrm{m}}-W_{\zeta\zeta}^{\mathrm{m}}}{2W_{\mathrm{m}}}
\:\!\right],
\end{align}
where we have used Eq.~\eqref{eq:ccond2}. 
Then we have
\begin{align}
E_{(2)(2)}^{\mathrm{m}}-E_{(1)(1)}^{\mathrm{m}}-\frac{G_{(0)(0)}^{\mathrm{m}}+G_{(3)(3)}^{\mathrm{m}}}{2}
=-\frac{e^{2\:\!(\psi_{\mathrm{m}}-\gamma_{\mathrm{m}})}}{W_{\mathrm{m}}^2}\left[\:\!
2W_{\mathrm{m}}^2\psi_{\rho\rho}^{\mathrm{m}}
+(W_{\rho}^{\mathrm{m}})^2
-W_{\mathrm{m}}W_{\rho\rho}^{\mathrm{m}}
\:\!\right]. 
\end{align}
Substituting this result into Eq.~\eqref{eq:V"m} and using the definition of $R$ in Eq.~\eqref{eq:R}, we immediately obtain $V''_{\mathrm{m}}$ in the form of Eq.~\eqref{eq:V"mEG}.


\begin{thebibliography}{99}

%
\bibitem{EventHorizonTelescope:2019dse}
K.~Akiyama \textit{et al.} (Event Horizon Telescope Collaboration),
Astrophys. J. Lett. \href{https://doi.org/10.3847/2041-8213/ab0ec7}{\textbf{875}, L1 (2019)}
[arXiv:\href{http://arxiv.org/abs/1906.11238}{1906.11238 [astro-ph.GA]}].

%
\bibitem{EventHorizonTelescope:2022wkp}
K.~Akiyama \textit{et al.} (Event Horizon Telescope Collaboration),
Astrophys. J. Lett. \href{https://doi.org/10.3847/2041-8213/ac6674}{\textbf{930}, L12 (2022)}
[arXiv:\href{http://arxiv.org/abs/2311.08680}{2311.08680 [astro-ph.HE]}].


%
\bibitem{LIGOScientific:2016aoc}
B.~P.~Abbott \textit{et al.} (LIGO Scientific Collaboration and Virgo Collaboration),
Phys. Rev. Lett. \href{https://doi.org/10.1103/PhysRevLett.116.061102}{\textbf{116}, 061102 (2016)}.

%
\bibitem{Perlick:2004}
V.~Perlick,
Living Rev. Relativity \href{https://doi.org/10.12942/lrr-2004-9}{\textbf{7}, 9 (2004)}.

%
\bibitem{Bozza:2002zj}
V.~Bozza,
Phys. Rev. D \href{https://doi.org/10.1103/PhysRevD.66.103001}{\textbf{66}, 103001 (2002)}
[arXiv:\href{https://arxiv.org/abs/gr-qc/0208075}{gr-qc/0208075}].

%
\bibitem{Tsukamoto:2016jzh}
N.~Tsukamoto,
Phys. Rev. D \href{https://doi.org/10.1103/PhysRevD.95.064035}{\textbf{95}, 064035 (2017)}
[arXiv:\href{https://arxiv.org/abs/1612.08251}{1612.08251 [gr-qc]}].



%
\bibitem{Shaikh:2019itn}
R.~Shaikh, P.~Banerjee, S.~Paul, and T.~Sarkar,
Phys. Rev. D \href{https://doi.org/10.1103/PhysRevD.99.104040}{\textbf{99}, 104040 (2019)}
[arXiv:\href{https://arxiv.org/abs/1903.08211}{1903.08211 [gr-qc]}].


%
\bibitem{Bozza:2001xd}
V.~Bozza, S.~Capozziello, G.~Iovane, and G.~Scarpetta,
Gen. Relativ. Gravit. \href{https://doi.org/10.1023/A:1012292927358}{\textbf{33}, 1535 (2001)}
[arXiv:\href{https://arxiv.org/abs/gr-qc/0102068}{gr-qc/0102068 [gr-qc]}].



%
\bibitem{Eiroa:2002mk}
E.~F.~Eiroa, G.~E.~Romero, and D.~F.~Torres,
Phys. Rev. D \href{https://doi.org/10.1103/PhysRevD.66.024010}{\textbf{66}, 024010 (2002)}
[arXiv:\href{https://arxiv.org/abs/gr-qc/0203049}{gr-qc/0203049}].

%
\bibitem{Tsukamoto:2016oca}
N.~Tsukamoto and Y.~Gong,
Phys. Rev. D \href{https://doi.org/10.1103/PhysRevD.95.064034}{\textbf{95}, 064034 (2017)}
[arXiv:\href{https://arxiv.org/abs/1612.08250}{1612.08250 [gr-qc]}].

%
\bibitem{Eiroa:2010wm}
E.~F.~Eiroa and C.~M.~Sendra,
Classical Quantum Gravity \href{https://doi.org/10.1088/0264-9381/28/8/085008}{\textbf{28}, 085008 (2011)}
[arXiv:\href{https://arxiv.org/abs/1011.2455}{1011.2455 [gr-qc]}].

\bibitem{Chen:2023uuy}
D.~Chen, Y.~Chen, P.~Wang, T.~Wu, and H.~Wu,
Eur. Phys. J. C \href{https://doi.org/10.1140/epjc/s10052-024-12950-z}{\textbf{84}, 584 (2024)}
[arXiv:\href{https://arxiv.org/abs/2309.00905}{2309.00905 [gr-qc]}].

%
\bibitem{Tsukamoto:2016qro}
N.~Tsukamoto,
Phys. Rev. D \href{https://doi.org/10.1103/PhysRevD.94.124001}{\textbf{94}, 124001 (2016)}
[arXiv:\href{https://arxiv.org/abs/1607.07022}{1607.07022 [gr-qc]}].

%
\bibitem{Nandi:2016uzg}
K.~K.~Nandi, R.~N.~Izmailov, A.~A.~Yanbekov, and A.~A.~Shayakhmetov,
Phys. Rev. D \href{https://doi.org/10.1103/PhysRevD.95.104011}{\textbf{95}, 
104011 (2017)}
[arXiv:\href{https://arxiv.org/abs/1611.03479}{1611.03479 [gr-qc]}].


%
\bibitem{Shaikh:2019jfr}
R.~Shaikh, P.~Banerjee, S.~Paul, and T.~Sarkar,
J. Cosmol. Astropart. Phys. \href{https://doi.org/10.1088/1475-7516/2019/07/028}{07, (2019) 028}; \href{https://doi.org/10.1088/1475-7516/2023/12/E01}{12 (2023) E01}.
[arXiv:\href{https://arxiv.org/abs/1905.06932}{1905.06932 [gr-qc]}].

%
\bibitem{Nandi:2006ds}
K.~K.~Nandi, Y.~Z.~Zhang, and A.~V.~Zakharov,
Phys. Rev. D \href{https://doi.org/10.1103/PhysRevD.74.024020}{\textbf{74}, 024020 (2006)}
[arXiv:\href{https://arxiv.org/abs/gr-qc/0602062}{gr-qc/0602062}].


%
\bibitem{TejeiroS:2005ltc}
J.~M.~Tejeiro S. and E.~A.~Larranaga R.,
Rom. J. Phys. \href{https://rjp.nipne.ro/2012_57_3-4.html}{\textbf{57}, 736 (2012)}
[arXiv:\href{https://arxiv.org/abs/gr-qc/0505054}{gr-qc/0505054}].

%
\bibitem{Bhattacharya:2019kkb}
A.~Bhattacharya and A.~A.~Potapov,
Mod. Phys. Lett. A \href{https://doi.org/10.1142/S0217732319500408}{\textbf{34}, 1950040 (2019)}.

%
\bibitem{Izmailov:2019tyq}
R.~N.~Izmailov, E.~R.~Zhdanov, A.~Bhattacharya, A.~A.~Potapov, and K.~K.~Nandi,
Eur. Phys. J. Plus \href{https://doi.org/10.1140/epjp/i2019-12914-5}{\textbf{134}, 384 (2019)}
[arXiv:\href{https://arxiv.org/abs/1909.13052}{1909.13052 [gr-qc]}].

%
\bibitem{Kubo:2016ada}
T.~Kubo and N.~Sakai,
Phys. Rev. D \href{https://doi.org/10.1103/PhysRevD.93.084051}{\textbf{93}, 084051 (2016)}.

%
\bibitem{Chakraborty:2016lxo}
S.~Chakraborty and S.~SenGupta,
J. Cosmol. Astropart. Phys. \href{https://doi.org/10.1088/1475-7516/2017/07/045}{07 (2017) 045}
[arXiv:\href{https://arxiv.org/abs/1611.06936}{1611.06936 [gr-qc]}].

%
\bibitem{Soares:2024rhp}
A.~R.~Soares, R.~L.~L.~Vit\'oria, and C.~F.~S.~Pereira,
Phys. Rev. D \href{https://doi.org/10.1103/PhysRevD.110.084004}{\textbf{110}, 084004 (2024)}
[arXiv:\href{https://arxiv.org/abs/2408.03217}{2408.03217 [gr-qc]}].

%
\bibitem{Nascimento:2020ime}
J.~R.~Nascimento, A.~Y.~Petrov, P.~J.~Porfirio, and A.~R.~Soares,
Phys. Rev. D \href{https://doi.org/10.1103/PhysRevD.102.044021}{\textbf{102}, 044021 (2020)}
[arXiv:\href{https://arxiv.org/abs/2005.13096}{2005.13096 [gr-qc]}].

%
\bibitem{Ishihara:2016sfv}
A.~Ishihara, Y.~Suzuki, T.~Ono, and H.~Asada,
Phys. Rev. D \href{https://doi.org/10.1103/PhysRevD.95.044017}{\textbf{95}, 044017 (2017)}
[arXiv:\href{https://arxiv.org/abs/1612.04044}{1612.04044 [gr-qc]}].

%
\bibitem{Takizawa:2021gdp}
K.~Takizawa and H.~Asada,
Phys. Rev. D \href{https://doi.org/10.1103/PhysRevD.103.104039}{\textbf{103}, 104039 (2021)}
[arXiv:\href{https://arxiv.org/abs/2103.10649}{2103.10649 [gr-qc]}].

%
\bibitem{Feleppa:2024kio}
F.~Feleppa, V.~Bozza, and O.~Y.~Tsupko,
Phys. Rev. D \href{https://doi.org/10.1103/PhysRevD.111.044018}{\textbf{111}, 044018 (2025)}
[arXiv:\href{https://arxiv.org/abs/2412.16712}{2412.16712 [gr-qc]}].

\bibitem{Gao:2025arj}
Y.~X.~Gao,
Eur. Phys. J. C \href{https://doi.org/10.1140/epjc/s10052-025-14528-9}{\textbf{85}, 794 (2025)}
[arXiv:\href{https://arxiv.org/abs/2503.06895}{2503.06895 [astro-ph.HE]}].

%
\bibitem{Sasaki:2025web}
T.~Sasaki,
Phys. Rev. D \href{https://doi.org/10.1103/7r5r-nlcj}{\textbf{112}, 024072 (2025)}
[arXiv:\href{https://arxiv.org/abs/2504.00355}{2504.00355 [gr-qc]}].


%
\bibitem{Bozza:2002af}
V.~Bozza,
Phys. Rev. D \href{https://doi.org/10.1103/PhysRevD.67.103006}{\textbf{67}, 103006 (2003)}
[arXiv:\href{https://arxiv.org/abs/gr-qc/0210109}{gr-qc/0210109}].

%
\bibitem{Dolan:2010wr}
S.~R.~Dolan,
Phys. Rev. D \href{https://doi.org/10.1103/PhysRevD.82.104003}{\textbf{82}, 104003 (2010)}
[arXiv:\href{https://arxiv.org/abs/1007.5097}{1007.5097 [gr-qc]}].


%
\bibitem{Hsieh:2021scb}
T.~Hsieh, D.~S.~Lee, and C.~Y.~Lin,
Phys. Rev. D \href{https://doi.org/10.1103/PhysRevD.103.104063}{\textbf{103}, 104063 (2021)}
[arXiv:\href{https://arxiv.org/abs/2101.09008}{2101.09008 [gr-qc]}].

%
\bibitem{AbhishekChowdhuri:2023ekr}
A.~Chowdhuri, S.~Ghosh, and A.~Bhattacharyya,
Front. Phys. \href{https://doi.org/10.3389/fphy.2023.1113909}{\textbf{11}, 1113909 (2023)}
[arXiv:\href{https://arxiv.org/abs/2303.02069}{2303.02069 [gr-qc]}].

%
\bibitem{Patil:2016oav}
M.~Patil, P.~Mishra, and D.~Narasimha,
Phys. Rev. D \href{https://doi.org/10.1103/PhysRevD.95.024026}{\textbf{95}, 024026 (2017)}
[arXiv:\href{https://arxiv.org/abs/1610.04863}{1610.04863 [gr-qc]}].

%
\bibitem{Chakrabarty:2022fbd}
H.~Chakrabarty and Y.~Tang,
Phys. Rev. D \href{https://doi.org/10.1103/PhysRevD.107.084020}{\textbf{107}, 084020 (2023)}
[arXiv:\href{https://arxiv.org/abs/2204.06807}{2204.06807 [gr-qc]}].

%
\bibitem{Igata:2025taz}
T.~Igata, 
arXiv:\href{https://arxiv.org/abs/2503.02320}{2503.02320 [gr-qc]}, to appear in \href{https://journals.aps.org/prd/accepted/10.1103/55vp-97gp}{Phys. Rev. D}.

%
\bibitem{Ferrari:1984zz}
V.~Ferrari and B.~Mashhoon,
Phys. Rev. D \href{https://doi.org/10.1103/PhysRevD.30.295}{\textbf{30}, 295 (1984)}.


%
\bibitem{Cardoso:2008bp}
V.~Cardoso, A.~S.~Miranda, E.~Berti, H.~Witek, and V.~T.~Zanchin,
Phys. Rev. D \href{https://doi.org/10.1103/PhysRevD.79.064016}{\textbf{79}, 064016 (2009)}
[arXiv:\href{https://arxiv.org/abs/0812.1806}{0812.1806 [hep-th]}].



%
\bibitem{Stefanov:2010xz}
I.~Z.~Stefanov, S.~S.~Yazadjiev, and G.~G.~Gyulchev,
Phys. Rev. Lett. \href{https://doi.org/10.1103/PhysRevLett.104.251103}{\textbf{104}, 251103 (2010)}
[arXiv:\href{https://arxiv.org/abs/1003.1609}{1003.1609 [gr-qc]}].

%
\bibitem{Raffaelli:2014ola}
B.~Raffaelli,
Gen. Relativ. Gravit. \href{https://doi.org/10.1007/s10714-016-2016-7}{\textbf{48}, 16 (2016)}
[arXiv:\href{https://arxiv.org/abs/1412.7333}{1412.7333 [gr-qc]}].







%
\bibitem{LIGOScientific:2016dsl}
B.~P.~Abbott \textit{et al.} (LIGO Scientific Collaboration and Virgo Collaboration),
Phys. Rev. Lett. \href{https://doi.org/10.1103/PhysRevLett.116.221101}{\textbf{116}, 221101 (2016)}
[arXiv:\href{https://arxiv.org/abs/1602.03841}{1602.03841 [gr-qc]}].

%
\bibitem{LIGOScientific:2020tif}
B.~P.~Abbott \textit{et al.} (LIGO Scientific Collaboration and Virgo Collaboration),
Phys. Rev. D \href{https://doi.org/10.1103/PhysRevD.103.122002}{\textbf{103}, 122002 (2021)}
[arXiv:\href{https://arxiv.org/abs/2106.15127}{2106.15127 [gr-qc]}].

%
\bibitem{Bolokhov:2025uxz}
S.~V.~Bolokhov and M.~Skvortsova,
Gravit. Cosmol. \href{https://link.springer.com/article/10.1134/S0202289325700306}{\textbf{31}, 423 (2025)}
[arXiv:\href{https://arxiv.org/abs/2504.05014}{2504.05014 [gr-qc]}].

%
\bibitem{Wald:1984}
R.~M.~Wald,
\textit{General Relativity} (University of Chicago Press, Chicago, 1984).

%
\bibitem{Griffiths:2012}
J.~B.~Griffiths and J.~Podolsk{\'y},
\textit{Exact Space-Times in Einstein's General Relativity} (Cambridge University Press, Cambridge, England, 2012).

%
\bibitem{Stephani:2003tm}
H.~Stephani, D.~Kramer, M.~A.~H.~MacCallum, C.~Hoenselaers, and E.~Herlt,
\textit{Exact Solutions of Einstein's Field Equations} (Cambridge University Press, Cambridge, England, 2003).

%
\bibitem{Koga:2019uqd}
Y.~Koga and T.~Harada,
Phys. Rev. D \href{https://doi.org/10.1103/PhysRevD.100.064040}{\textbf{100}, 064040 (2019)}
[arXiv:\href{https://arxiv.org/abs/1907.07336}{1907.07336 [gr-qc]}].

%
\bibitem{Assumpcao:2018bka}
T.~Assumpcao, V.~Cardoso, A.~Ishibashi, M.~Richartz, and M.~Zilhao,
Phys. Rev. D \href{https://doi.org/10.1103/PhysRevD.98.064036}{\textbf{98}, 064036 (2018)}
[arXiv:\href{https://arxiv.org/abs/1806.07909}{1806.07909 [gr-qc]}].

%
\bibitem{Shipley:2016omi}
J.~Shipley and S.~R.~Dolan,
Classical Quantum Gravity \href{https://doi.org/10.1088/0264-9381/33/17/175001}{\textbf{33}, 175001 (2016)}
[arXiv:\href{https://arxiv.org/abs/1603.04469}{1603.04469 [gr-qc]}].


%
\bibitem{Nakashi:2019mvs}
K.~Nakashi and T.~Igata,
Phys. Rev. D \href{https://doi.org/10.1103/PhysRevD.99.124033}{\textbf{99}, 124033 (2019)}
[arXiv:\href{https://arxiv.org/abs/1903.10121}{1903.10121 [gr-qc]}].


%
\bibitem{Allahyari:2018cmg}
A.~Allahyari, H.~Firouzjahi, and B.~Mashhoon,
Phys. Rev. D \href{https://doi.org/10.1103/PhysRevD.99.044005}{\textbf{99}, 044005 (2019)}
[arXiv:\href{https://arxiv.org/abs/1812.03376}{1812.03376 [gr-qc]}].



%
\bibitem{Pradhan:2012rkk}
P.~Pradhan,
Pramana \href{https://doi.org/10.1007/s12043-016-1214-x}{\textbf{87}, 5 (2016)}
[arXiv:\href{https://arxiv.org/abs/1205.5656}{1205.5656 [gr-qc]}].


%
\bibitem{Virbhadra:1999nm}
K.~S.~Virbhadra and G.~F.~R.~Ellis,
Phys. Rev. D \href{https://doi.org/10.1103/PhysRevD.62.084003}{\textbf{62}, 084003 (2000)}
[arXiv:\href{https://arxiv.org/abs/astro-ph/9904193}{astro-ph/9904193 [astro-ph]}].

%
\bibitem{Maldacena:2015waa}
J.~Maldacena, S.~H.~Shenker, and D.~Stanford,
J. High Energy Phys. \href{https://doi.org/10.1007/JHEP08(2016)106}{\textbf{08} (2016) 106}
[arXiv:\href{https://arxiv.org/abs/1503.01409}{1503.01409 [hep-th]}].

%
\bibitem{Hashimoto:2016dfz}
K.~Hashimoto and N.~Tanahashi,
Phys. Rev. D \href{https://doi.org/10.1103/PhysRevD.95.024007}{\textbf{95}, 024007 (2017)}
[arXiv:\href{https://arxiv.org/abs/1610.06070}{1610.06070 [hep-th]}].


%
\bibitem{Giataganas:2024hil}
D.~Giataganas, A.~Kehagias, and A.~Riotto,
J. High Energy Phys. \href{https://doi.org/10.1007/JHEP09(2024)168}{09 (2024) 168}
[arXiv:\href{https://arxiv.org/abs/2403.10605}{2403.10605 [gr-qc]}].

%
\bibitem{Gallo:2024wju}
E.~Gallo and T.~M\"adler,
Eur. Phys. J. C \href{https://doi.org/10.1140/epjc/s10052-025-14046-8}{\textbf{85}, 299 (2025)}
[arXiv:\href{https://arxiv.org/abs/2412.10328}{2412.10328 [gr-qc]}].







\end{thebibliography}
\end{document}